\documentclass[sigconf]{acmart}

\usepackage{microtype}
\newcommand{\tighttt}[1]{%
  {\small\texttt{#1}}%
}
\usepackage{xcolor}
\usepackage{enumitem}
\usepackage{array}

\definecolor{inputcolor}{HTML}{1F77B4}   % Blue
\definecolor{outputcolor}{HTML}{D62728}  % Red

\newcommand{\name}{{\tt StateScribe}}

\usepackage{comment}
\usepackage{framed}
\usepackage{ifthen}
\usepackage{xspace}
\usepackage{etoolbox}

\AtBeginDocument{%
  }

\copyrightyear{2026}
\acmYear{2026}
\setcopyright{cc}
\setcctype{by}
\acmConference[UIST '26]{The 39th Annual ACM Symposium on User Interface Software and Technology}{November 02--05, 2026}{Detroit, MI, USA}
\acmBooktitle{The 39th Annual ACM Symposium on User Interface Software and Technology (UIST '26), November 02--05, 2026, Detroit, MI, USA}
\acmDOI{10.1145/3830398.3830475}
\acmISBN{979-8-4007-2856-3/2026/11}

\def\name{StateScribe}
\begin{document}

\title{{\name}: Towards Accessible Change Awareness Across Real-World Revisits}
\author{Ruei-Che Chang}
\authornote{Both authors contributed equally to this work.}
\affiliation{
  \institution{University of Michigan}
  \city{Ann Arbor, MI}
  \country{USA}
}
\email{rueiche@umich.edu}

\author{Xirui Jiang}
\authornotemark[1]
\affiliation{
  \institution{University of Michigan}
  \city{Ann Arbor, MI}
  \country{USA}
}
\email{xirui@umich.edu}

\author{Rosiana Natalie}
\affiliation{
 \institution{University of Michigan}
 \city{Ann Arbor, MI}
 \country{USA}
}
\email{rosianan@umich.edu}

\author{Hao Chen}
\affiliation{
 \institution{Samsung Research America}
 \city{Plano, TX}
 \country{USA}
}
\email{hao.chen1@samsung.com}

\author{Vlad Roznyatovskiy}
\affiliation{
 \institution{Samsung Research America}
 \city{Plano, TX}
 \country{USA}
}
\email{vlad.r@samsung.com}

\author{Jianzhong Zhang}
\affiliation{
 \institution{Samsung Research America}
 \city{Plano, TX}
 \country{USA}
}
\email{jianzhong.z@samsung.com}

\author{Kang G. Shin}
\affiliation{
 \institution{University of Michigan}
 \city{Ann Arbor, MI}
 \country{USA}
}
\email{kgshin@umich.edu}

\author{Ke Sun}
\affiliation{
 \institution{University of Michigan}
 \city{Ann Arbor, MI}
 \country{USA}
}
\email{kesuniot@umich.edu}

\author{Anhong Guo}
\affiliation{
 \institution{University of Michigan}
 \city{Ann Arbor, MI}
 \country{USA}
}
\email{anhong@umich.edu}

\begin{teaserfigure}
\vspace{-1.2pc}
  \includegraphics[width=\textwidth, alt={}]{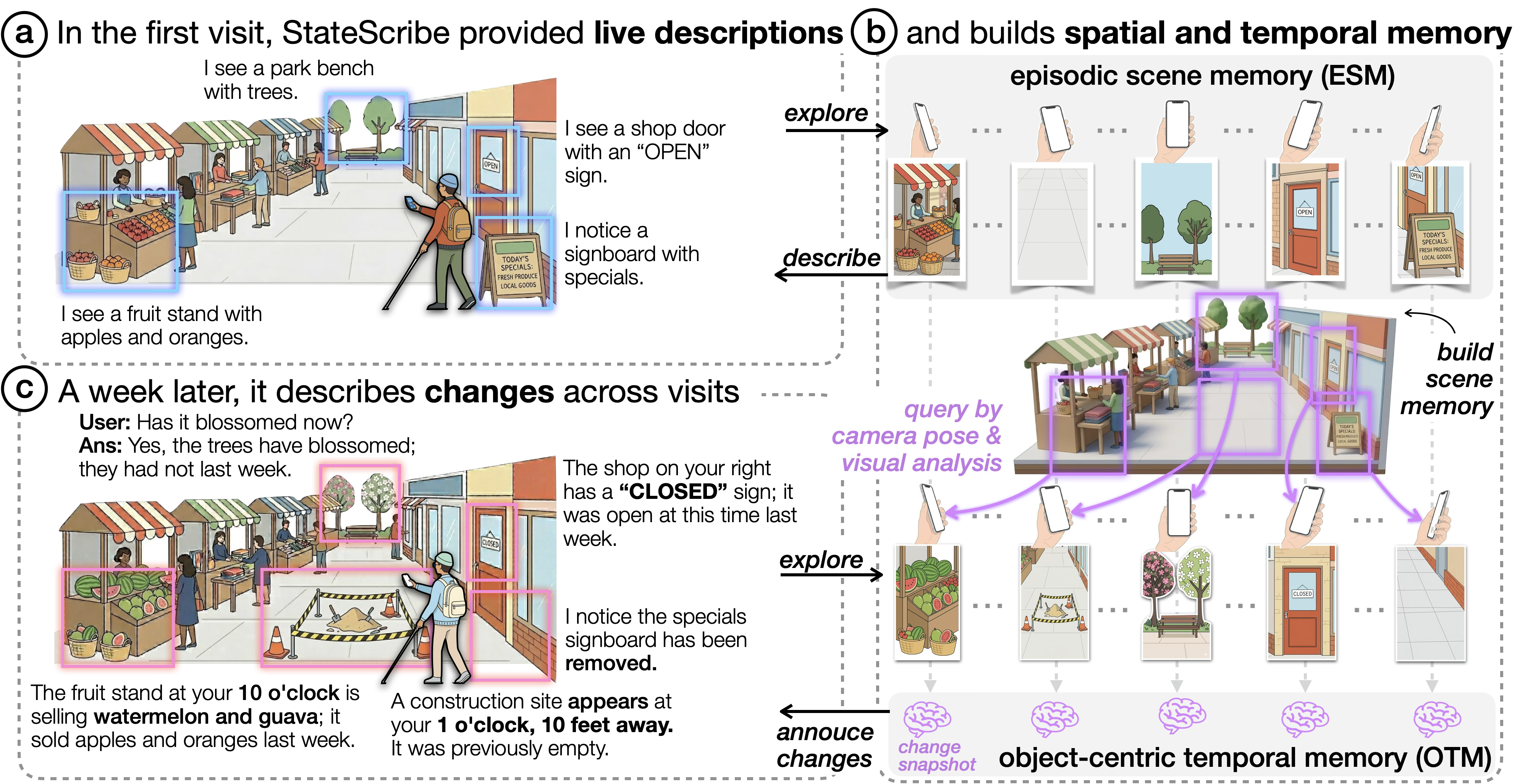}
  \vspace{-1.5pc}
  \caption{
  {\name} enables accessible change awareness across real-world revisits. 
  When visiting a new location, (a) {\name} provides live descriptions during 
  exploration to support scene understanding. Simultaneously, (b) it builds spatial 
  and temporal memory through a novel dual-layer memory architecture, including an 
  Episodic Scene Memory (ESM), which stores recent scene observations, and an 
  Object-Centric Temporal Memory (OTM), which tracks object-level changes over time. Upon 
  revisiting the same location, (c) {\name} retrieves and compares past views via camera pose and visual similarity, and announces changes that have occurred since the previous visit, including object appearance, removal, replacement, or content modifications.
  }
  \Description{Figure 1
The figure is composed of three panels labeled (a), (b), and (c), illustrating how a system called ``StateScribe'' provides live descriptions, builds memory, and reports changes across visits.
In panel (a), titled In the first visit, StateScribe provided live descriptions,'' a street market scene is shown with a person using a white cane, indicating a visually impaired user. Several regions of the scene are highlighted with blue boxes. Text descriptions are placed near these highlights, including: I see a fruit stand with apples and oranges,'' I see a park bench with trees,'' I see a shop door with an ‘OPEN’ sign,'' and I notice a signboard with specials.'' The scene includes market stalls with striped awnings, baskets of fruit, a bench with trees in the background, and a storefront with a visible OPEN'' sign. These descriptions demonstrate how the system narrates key elements of the environment during the first visit.
In panel (b), titled and builds spatial and temporal memory,'' the diagram shows a sequence labeled episodic scene memory (ESM).'' Small image snapshots of different parts of the scene (e.g., fruit stand, empty walkway, bench with trees, shop door, and signboard) are arranged in order, connected by arrows labeled explore'' and describe.'' A larger reconstructed scene appears below, with purple outlines marking regions corresponding to the stored snapshots. Arrows labeled query by camera pose & visual analysis'' connect the stored snapshots to the reconstructed scene, and a label build scene memory'' indicates the integration process. At the bottom, a sequence labeled object-centric temporal memory (OTM)'' shows individual objects (such as fruit, construction cones, trees, and a door with a sign) tracked over time, with brain icons and the label change snapshot.'' This panel illustrates how the system builds both spatial memory of the scene and temporal memory of objects.
In panel (c), titled A week later, it describes changes across visits,'' a similar market scene is shown with pink highlights indicating differences from the previous visit. A dialogue appears at the top: User: Has it blossomed now?'' and Ans: Yes, the trees have blossomed; they had not last week.'' Additional textual descriptions identify changes: The shop on your right has a ‘CLOSED’ sign; it was open at this time last week,'' I notice the specials signboard has been removed,'' The fruit stand at your 10 o’clock is selling watermelon and guava; it sold apples and oranges last week,'' and A construction site appears at your 1 o’clock, 10 feet away. It was previously empty.'' These changes are visually indicated by highlighted regions such as blossoming trees, a closed shop door, a modified fruit stand, and a new construction area. Arrows labeled explore'' and ``announce changes'' connect to the memory representations below, showing how the system detects and communicates differences over time.
}
  \label{fig:teaser}
  \vspace{0pc}
\end{teaserfigure}

\begin{abstract}
Real-world environments evolve continuously, yet blind and low-vision (BLV) individuals often have limited access to understanding how they change over time. 
Unexpected or relocated objects, layout modifications, and content updates 
(e.g., price changes) can introduce safety risks and cognitive burden. 
While existing visual assistive technologies can describe immediate surroundings, they operate as one-off interactions and lack mechanisms to surface meaningful changes across revisits.
Informed by a survey of 33 BLV individuals, we develop {\name}, a system that supports accessible awareness of real-world changes across revisits. {\name} employs a dual-layer memory architecture that integrates episodic scene memory and object-centric temporal memory to enable scalable and structured change tracking. 
It provides both live descriptions of the current scene, and descriptions of 
\textit{what} has changed, \textit{when} and \textit{where} it occurred across revisits, such as \emph{``The shop on your right has a “CLOSED” sign; it was open at this time last week.''}
Our evaluation shows that {\name} maintains high accuracy ($F_1$-score = \textbf{83.1\%}) across 11 revisits, while remaining low-latency (mean\textbf{<1.42s}) and memory-efficient (\textbf{<55MB}) across 110 revisits.
A user study with nine BLV participants demonstrates that {\name} improves change awareness across revisits in three real-world locations. 
Finally, we discuss implications for long-term AI-assisted companions that support broader change observation using multimodal sensing, extend beyond changes to other memory capabilities, and adapt to individual users, intents, and contexts.
\end{abstract}



% \settopmatter{printfolios=true}

\maketitle 
\renewcommand{\shortauthors}{Chang and Jiang et al.}

\section{Introduction}
Real-world environments evolve continuously with many unanticipated changes,
yet blind and low-vision (BLV) individuals often have limited access to information that helps them understand how familiar spaces transform over time. 
For example, newly introduced obstacles, relocated objects in shared spaces, content changes (\textit{e.g.,} price updates), or layout modifications can pose safety risks and increase cognitive burden. 
Despite these challenges, there remains a limited understanding of what constitutes meaningful real-world changes for BLV individuals, and how to convey them effectively.

While existing human-powered services (\textit{e.g.,} Be My Eyes~\cite{bemyeyes}, Aira~\cite{aira}), AI-powered visual assistive technologies (\textit{e.g.,} Seeing AI~\cite{seeingai}, Be My AI~\cite{bemyai}, WorldScribe~\cite{WorldScribe}), or live video AI (\textit{e.g.,} ChatGPT Live Video~\cite{gpt_live}, Gemini Live~\cite{gemini_live}) can describe immediate surroundings, they typically operate as one-off sessions that retain only within-session memory. 
As a result, they lack persistent 3D spatial memory of a location~\cite{Chang2025probing} and mechanisms for surfacing meaningful changes across revisits.
Enabling such longitudinal assistance poses several challenges: 
\emph{(i)} \textbf{Maintaining longitudinal memory in 3D space is difficult}, as systems must identify and store the most relevant observations from the 3D real world without overloading with redundant information. 
\emph{(ii)} \textbf{Scaling memory storage and retrieval over extended use is difficult}, as spatial and temporal data grow continuously, increasing retrieval latency and confusion. 
\emph{(iii)} \textbf{Incorporating change-awareness into real-time interactions can easily overwhelm users}, especially if updates are frequent and noisy. 

Based on a survey of 33 BLV individuals, we identified categories of changes that are difficult to perceive yet significantly affect daily life, such as newly introduced obstacles in shared spaces (\textit{e.g.,} scattered packages), changes to signage along routine commuting routes (\textit{e.g.,} removed or added signage), and updated pricing or layout during grocery shopping. 
To address these, we present {\name}, a system that supports accessible awareness of real-world changes through a smartphone and its sensor data. 
{\name} employs a dual-layer memory architecture that integrates Episodic Scene Memory (ESM) and Object-centric Temporal Memory (OTM). 
Specifically, ESM continuously updates information from recent visits within a context window, archiving older data to maintain an up-to-date representation of the scene. 
In contrast, OTM selectively maintains key objects, tracking their states and spatial locations across visits while recording meaningful changes. 
This design enables scalable, structured, and efficient storage and retrieval.

Using these memory structures and smartphone 3D data (\textit{e.g.,} RGB-D images and camera 
intrinsics), {\name} retrieves ESM and OTM by querying the current camera pose to 
identify the most recent frames from prior visits and detect significant changes.
It generates descriptions that clearly 
communicate \textit{\textbf{what has changed, and when and where the change occurred.}} 
For example, a user revisiting the market (Figure~\ref{fig:teaser}b) a week later might hear: \textit{``The fruit stand at your 10 o'clock is selling watermelon and guava; it sold apples and oranges last week,''} \textit{``The bakery on your right now shows a `CLOSED' sign; it was open at this time last week,''} or \textit{``A construction site is at your 1 o'clock, 10 feet away.''} 
Beyond change descriptions, {\name} also provides live descriptions (Figure~\ref{fig:teaser}a) to support real-time scene understanding, using a similar but simplified approach to WorldScribe~\cite{WorldScribe}. 
{\name} further enables interactive exploration of how their environment has evolved, where users can query via speech, \textit{e.g.,} \textit{``Has it blossomed now?''}, with responses: \textit{``Yes, the trees have blossomed; they had not last week''}(Figure~\ref{fig:teaser}b).

In a technical evaluation against two baselines, a \textit{Live model} that actively prompts for change detection and a \textit{Offline video model} that processes full-visit videos, {\name} achieved an $F_1$-score of \textbf{83.1\%} across \textbf{11} visits (291 changes), outperforming the \textit{Live model} (40.1\%) and \textit{Offline video model} (27.4\%). It also showed higher spatial accuracy, with lower clock-direction errors (in hour units; \textbf{M=0.24} vs. 1.98 and 1.86) and smaller distance estimation errors (in feet; \textbf{M=0.68} vs. 5.29 and 6.49). Additionally, {\name} remained low-latency (\textbf{mean<1.42s}) and memory-efficient (\textbf{<55MB}) across \textbf{110} revisits in simulated long-term use.
A user study with 9 BLV participants showed that {\name} enhanced change awareness across revisits in different settings (\textit{e.g.,} office, grocery store, courtyard). 

Finally, we discuss extending {\name} to video-based assistive technologies, including remote sighted assistance (RSA) services~\cite{aira, bemyeyes} and live video AI systems~\cite{gemini_live, gpt_live}, and derive design implications for multimodal, personalized, long-term AI companions with broader memory capabilities that adapt to users, intents, and contexts.

\section{Related Work}
{\name} builds on prior work to provide BLV people with visual 
descriptions for real-world understanding, as well as traditional and 
AI-enabled memory-augmented systems. In what follows, we describe 
our motivation from prior related work and research gaps.

\subsection{Visual Descriptions for BLV People}
Textual descriptions are essential for making digital and physical 
information accessible to BLV people. 
Recent advances in VLMs enabled automated image descriptions and 
conversational interaction, reducing reliance on human-authored alt text. 
These capabilities allowed BLV users to quickly obtain visual summaries 
and interactively explore details, which are increasingly integrated 
into social network platforms (\textit{e.g.,} Facebook~\cite{Wu2017Facebook}) and 
assistive apps (\textit{e.g.,} Seeing AI~\cite{seeingai}, Be My AI~\cite{bemyai}).
In addition, prior work has explored integrating VLMs into systems that 
provided continuous, live visual descriptions. Such systems dynamically 
adapt the level and content of information based on user behavior, 
including camera motion~\cite{WorldScribe} and hand 
movements~\cite{TouchScribe}.

Beyond visual descriptions, prior work in computer vision developed 
models for detecting and summarizing visual changes across 
images~\cite{jhamtani2018learning, forbes2019neural}. In parallel, 
HCI researchers explored systems that describe visual differences
between image pairs to support an accessible collaborative slide 
editing~\cite{Diffscriber}, image editing
\cite{VizXpress, EditScribe, GenAssist}, and graphics and 3D geometry 
editing~\cite{A11yBoard, A11yShape}.
However, support for detecting and describing real-world visual changes 
remains limited. Environments change continuously and unpredictably, 
which can impact BLV individuals’ daily lives.
For instance, unexpected barriers in outdoor 
environments~\cite{cushley2023unseen, Williams2023, el2023survey}, 
relocated objects in indoor shared spaces~\cite{turkstra2025assistive, 
patil2025designing}, or seasonal product rearrangements and price updates 
in retail environments~\cite{Lee2021Product, khattab2015understanding, 
yu2015retail, tullio2021you} could potentially introduce safety risks 
and cognitive burdens for BLV individuals.
To close these gaps, we first categorized meaningful changes through a 
formative study and developed {\name}, a system that possesses 
spatial and temporal memory to detect and describe meaningful changes 
across visits on familiar routes or locations, beyond live descriptions 
of immediate surroundings. 

\subsection{Human Memory-Augmented Interactions}
Early work in human memory-augmented interactions~\cite{lamming1994design} 
took the form of lifelogging systems, enabling applications such as 
social or audio reminders~\cite{hayes2004personal}, contextual 
suggestions~\cite{rhodes1997wearable}, sharing everyday 
moments~\cite{hayes2004personal}, and support for people with memory 
impairments~\cite{Lee2007, Lee2008, hodges2006sensecam}. While effective, 
these systems faced key challenges related to capturing meaningful data, 
managing storage constraints, and enabling timely retrieval of past data.

%DDDDD
First, they typically relied on continuously capturing large portions of everyday experiences across modalities, including photos~\cite{hodges2006sensecam, Gemmell2004, mann2005designing, Footprint}, videos~\cite{mann1998wearcam, sawahata2003wearable, Hori2003}, audio signals~\cite{rhodes1997wearable, vemuri2004audio, hayes2004personal, Memoro}, GPS data~\cite{Gemmell2004, Hori2003, rhodes1997wearable, Footprint}, and physiological signals~\cite{Hori2003, healey1998startlecam, Chan2020Biosignal, Memento, Prospero, Prompto}. 
However, such continuous recording imposed substantial storage demands, limiting long-term usability. To mitigate this, some systems adopted selective capture strategies, such as manual recording~\cite{lamming1994design}, biosignal-based triggers (\textit{e.g.,} startle responses~\cite{healey1998startlecam} or emotional changes~\cite{Memento}), and spatial triggers~\cite{Footprint} (\textit{e.g.,} remaining within a 50-meter radius for over five minutes).
Beyond storage, accessing and making use of captured data remained challenging. Traditional interfaces primarily supported browsing by time~\cite{Cooper2005}, location~\cite{toyama2003geographic}, or both~\cite{chen2006browsing, Gemmell2004, Footprint}, offering limited support for efficient browsing and retrieval of relevant past information.

Recent advances in multimodal large language models (MLLMs) and retrieval-augmented generation (RAG) have improved how memory is captured, stored, and retrieved.
Prior computer vision work explored memory structures for embodied agents across tasks such as spatial reasoning~\cite{zhu2025struct2d}, robot navigation~\cite{hu20253dllmm}, and persistent 3D scene understanding~\cite{fan2025embodied}, but these were optimized for offline accuracy rather than real-time human-AI interaction.
In contrast, systems like Memoro~\cite{Memoro} store conversational snippets as lightweight memory units for efficient live retrieval, while OmniQuery~\cite{omniquery} extracts semantic signals from images and videos for structured indexing. Memory Reviver further organizes photo collections into hierarchical representations to support conversational retrieval for BLV users~\cite{MemoryReviver}. However, these largely text-based approaches remain limited in dynamic real-world settings, where rich visual information cannot be fully captured through text alone.
% Moreover, over the long term, the volume of textual data can rapidly exceed what LLMs can effectively process~\cite{xu2021beyond, liu2024lost}.

To address this, {\name} employs a novel memory architecture that captures visual information from camera frames and selectively retains meaningful object changes in both visual and textual modalities across recent visits, while archiving outdated data to maintain lightweight storage for extended use. 
{\name} also has an efficient memory storage and retrieval pipeline that leverages camera poses and visual analysis, enabling it to proactively announce changes in real-time and across visits.

\begin{figure*}[t]
\vspace{-1pc}
\begin{center}
\includegraphics[width=\linewidth]{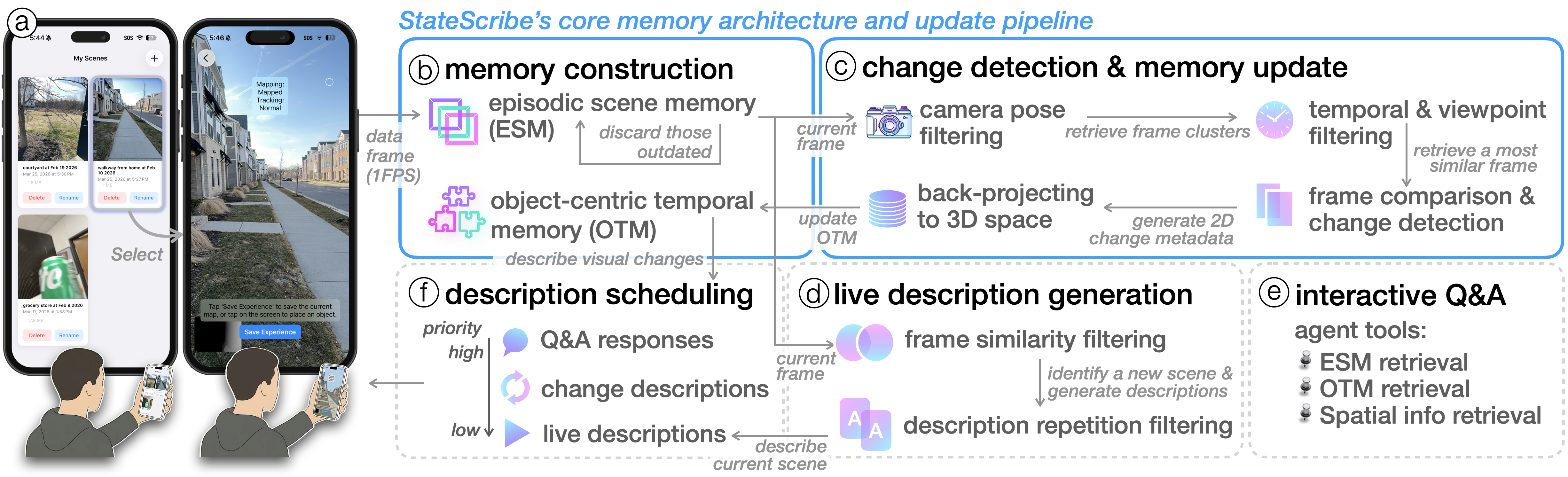}
\vspace{-1.8pc}
\caption{\name's system architecture and processing pipeline. 
(a) {\name}'s mobile interfaces.
(b) {\name} constructs episodic scene memory (ESM) from streaming frames 
and updates object-centric temporal memory (OTM) through a change-detection pipeline. 
(c) For each incoming data frame (consisting of RGB-D images, camera intrinsics), 
{\name} retrieves the most relevant reference frame and compares it to generate 
change metadata, which is then used to update the OTM. 
(d) In parallel, ESM supports live scene description by filtering visually 
similar scenes and avoiding repetitive descriptions. 
(e) Users can ask questions about the current scene, past scenes, or detected changes; 
a Q\&A agent answers these queries using built-in tools such as object 
localization and retrieval of recent scenes or changes. 
(f) Descriptions are scheduled and delivered based on priority: Q\&A responses, 
change descriptions, and then live scene descriptions.
}
\label{fig:pipeline}
\Description{Figure 2
The figure presents an overview of ``StateScribe’s core memory architecture and update pipeline,'' organized into multiple labeled components (a) through (f), showing how visual data is processed, stored, and used for descriptions and interaction.
In panel (a), two smartphone screens are shown under the label My Scenes.'' The left phone displays a gallery of captured scenes with thumbnail images and buttons such as Delete'' and Resume.'' The right phone shows a live camera view of a street with buildings and a walkway, along with interface text including Mapping, Tracking: Normal'' and a button labeled Save Experience.'' Below the phones, a user is depicted holding the device, indicating scene capture. An arrow labeled data frame (1FPS)'' points from this panel into the system pipeline, indicating that image frames are continuously collected.
In panel (b), titled memory construction,'' two types of memory are shown: episodic scene memory (ESM)'' and object-centric temporal memory (OTM).'' The ESM component includes a note discard those outdated,'' indicating that older or irrelevant frames are filtered out. The OTM component is labeled ``describe visual changes,'' indicating that it tracks how objects evolve over time. Both memory types receive input from the incoming data frames and contribute to subsequent processing.
In panel (c), titled change detection \& memory update,'' a sequence of processing steps is shown. Starting from a current frame,'' the pipeline includes camera pose filtering,'' followed by temporal & viewpoint filtering'' with a note retrieve a most similar frame.'' The process continues with frame comparison & change detection.'' In parallel, there is a step labeled back-projecting to 3D space,'' followed by generate 2D change metadata,'' which is used to ``update OTM.'' These steps illustrate how the system identifies differences between frames and updates its memory structures accordingly.
In panel (d), titled live description generation,'' the system processes the current frame'' using frame similarity filtering'' to identify a new scene & generate descriptions.'' It also applies ``description repetition filtering'' to avoid redundant outputs. This component produces real-time textual descriptions of the scene.
In panel (e), titled interactive Q\&A,'' a list of agent tools is shown, including ESM retrieval,'' OTM retrieval,'' and Spatial info retrieval.'' These tools support user queries by accessing different types of stored information.
In panel (f), titled description scheduling,'' outputs are prioritized along a vertical scale from high'' to low.'' The categories include Q&A responses'' (highest priority), change descriptions,'' and live descriptions'' (lowest priority). An arrow labeled ``describe current scene'' indicates that the system selects which type of description to produce based on priority.
}
\vspace{-1pc}
\end{center}
\end{figure*}

\section{Formative Study: Understanding Real-World Changes Meaningful to BLV People}
To address the key question, \textit{What constitutes meaningful real-world changes for BLV people?} We surveyed 33 participants (16 blind, 17 low vision; 10 men, 21 women, 2 non-binary; mean age = 45.1, SD = 15.9) recruited through local organizations and prior contacts (Table~\ref{tab:demographic-formative}). We analyzed responses using affinity diagramming.

\subsection{Results}

\subsubsection{Changes during daily commutes}
Participants frequently reported \textbf{temporary barriers} near construction sites (N=28), such as fences, traffic cones, scaffolding, tape, and detours, which obstructed their usual routes. The construction sites also sometimes \textbf{removed or obscured key landmarks} (N=7), including signage, pedestrian signals, and tactile paving. Also, natural changes, such as snowbanks, flooding (N=23), and fallen or overhanging tree branches, further disrupted familiar routes.
Furthermore, the \textbf{replacement of familiar landmarks} caused confusion, as F15 noted: \textit{``A bus stop that used to be just a bench was replaced with a metal shelter. I didn’t expect it and bumped into it.''} While some high-level changes can be anticipated (\textit{e.g.,} via weather or transit updates), pedestrian-level changes remained difficult to detect, with little warning (F20: \textit{``no commotion or construction noise''}). Participants often noticed them only after collisions, cane detection, or others’ alerts.

\subsubsection{Changes during grocery shopping}
Beyond \textbf{temporary barriers} (\textit{e.g.,} carts or displays blocking aisles), 
participants reported \textbf{spatial changes} during grocery shopping, 
including product relocations (N=24) and aisle rearrangements (N=21). 
These unexpected changes caused frustration (F18: \textit{``The chip aisle 
became canned food that I had to find someone to locate it.''}).
Also, \textbf{price changes and sales signage} (N=19) often went unnoticed, 
leading to surprise at checkout (F6: \textit{``I defaulted to the old price.''}). 
To cope with these, some used RSA services (\textit{e.g.,} 
Aira~\cite{aira}) or AI-powered apps (\textit{e.g.,} Seeing AI~\cite{seeingai}), 
occasionally scanning the environment to locate items (F28).
However, most relied on nearby shoppers or store staff and expressed 
a desire for advance awareness of such changes.

\subsubsection{Changes within the home}
Participants frequently encountered \textbf{temporary obstacles or clutter} 
(N=26), including blocked pathways (N=16), furniture rearrangements (N=11),
and seasonal decorations (N=9). 
Such changes stemmed from both participants (N=9) and other household members, 
with most living with family (N=14) or roommates (N=7). 
For example, F20 noted that everyday household behaviors (\textit{e.g.,} leaving boxes 
on the floor) introduced disruptive clutter.
These disruptions were particularly challenging as many participants did 
not use a cane in familiar home environments, leading to collisions 
(F28: \textit{``A chair left in my walking path caused me to fall.''}).
While a few low-vision participants could detect changes under good lighting,
most became aware only through physical contact, often resulting in 
frustration or minor injuries.

\subsection{Summary of Meaningful Visual Changes}\label{3_change_category}
In sum, participants mainly relied on memory or direct encounters to navigate 
familiar environments, occasionally using assistive tools (\textit{e.g.,} white canes, 
mobile AT) to manage unexpected changes. However, awareness remained largely reactive, 
highlighting the need for proactive information and advance alerts to reduce mobility difficulties, hazards, and frustration. Based on these insights, we 
categorize meaningful changes as follows:

\begin{itemize}[leftmargin=1.2em]
\item[\textit{(i)}] \textbf{Object-Level Changes:} Changes involving the presence 
  or the identity of individual objects at a location.
\begin{itemize}[leftmargin=1em]
    \item \textbf{Appeared:} Newly introduced objects not previously present 
       (\textit{e.g.,} construction barriers, clutter).
    \item \textbf{Removed:} Previously existing objects that are no longer 
       present (\textit{e.g.,} missing landmarks or signage).
    \item \textbf{Replaced:} Objects substituted by different ones at the 
      same location (\textit{e.g.,} a bench replaced by a bus shelter).
\end{itemize}

\item[\textit{(ii)}] \textbf{Spatial Changes:} Changes that alter the arrangement 
   or navigability of the environment.
\begin{itemize}[leftmargin=1em]
    \item \textbf{Relocated:} Objects moved to new locations (\textit{e.g.,} product 
       relocations in stores).
    \item \textbf{Layout Changes:} Modifications to spatial layout that affect mobility (\textit{e.g.,} aisle rearrangements, blocked pathways).
\end{itemize}
\item[\textit{(iii)}] \textbf{Attribute Changes:} Changes in the properties or state 
of existing objects without altering their identity or position (\textit{e.g.,} price changes, 
signage updates, on/off status). 

\end{itemize}

\section{{\name}}
{\name}\footnote{Code and datasets are available at \href{https://github.com/HumanAILab/StateScribe}{https://github.com/HumanAILab/StateScribe}} is a system with spatial and temporal memory that observes and 
describes real-world changes across visits, going beyond live descriptions 
of the current scene. 
For example, when a BLV user explores a new market (Figure~\ref{fig:teaser}),
{\name} provides live descriptions for scene understanding, such as 
\textit{``I see a fruit stand with apples and oranges''} or \textit{``I see 
a bakery with an `OPEN' sign on its door.''} 
Meanwhile, it continuously builds a memory from captured frames.
Upon revisiting the market a week later, {\name} compares the current view 
with stored memory and announces changes during exploration, \textit{e.g.,} 
\textit{``The fruit stand at your 10 o'clock is selling watermelon and guava; 
it sold apples and oranges last week,''} \textit{``The bakery on your right 
now shows a `CLOSED' sign; it was open at this time last week,''} 
or \textit{``A construction site is at your 1 o'clock, 10 feet away.''} 
Users can also query temporal changes via speech, \textit{e.g.,} \textit{``Has it
blossomed now?''}, receiving responses such as \textit{``Yes, the trees have 
blossomed; they had not last week.''}
Below, we describe how {\name} achieves this experience with different 
system modules. 

\subsection{Memory Construction Module}\label{4_memory_architecture}
In this module, {\name} streams essential sensor data from the smartphone 
to the backend for processing and storage within a novel dual-layer memory 
architecture: \textit{(i)} Episodic Scene Memory and \textit{(ii)} Object-Centric Temporal Memory.
These memory structures are informed by our formative study, which shows that 
observed changes primarily fall into three categories: single-object changes,
spatial changes (\textit{e.g.,} multi-object rearrangement), and attribute changes within
an object. 
However, this design may be limited in capturing fast motion or activity-based 
changes, which may require additional sensors and computational resources. 
We leave these as future work and discuss in Section~\ref{7_discussion}.

\textbf{Episodic Scene Memory (ESM).}
To capture the user’s visual context in real time,
we design ESM to retain full observations within a configurable temporal 
context window (\textit{e.g.,} the last 10 days, 5 hours, or single visit), 
while archiving older data (Figure~\ref{fig:pipeline}a). 
ESM serves as the primary memory retrieval source, 
storing data streamed from the mobile app to the backend at one data 
frame per second (FPS), including RGB-D frames, confidence maps, camera poses, and associated metadata. Upon receipt, the backend resizes images, 
refines the depth maps using confidence maps to filter out low-confidence pixels, generates visual embeddings, and asynchronously compresses and stores each instance as an \textit{ESM data frame} on local disk.
We adopt a frame-based representation over 3D point clouds because keyframes provide higher visual quality for VLM-based understanding and can later be compared directly for change detection (Section~\ref{4_change_detection}). Spatial information can be encoded efficiently as text using camera poses and 3D bounding boxes, enabling faster, more memory-efficient comparison and retrieval.

\begin{figure}[t!]
\centering
\includegraphics[width=\linewidth]{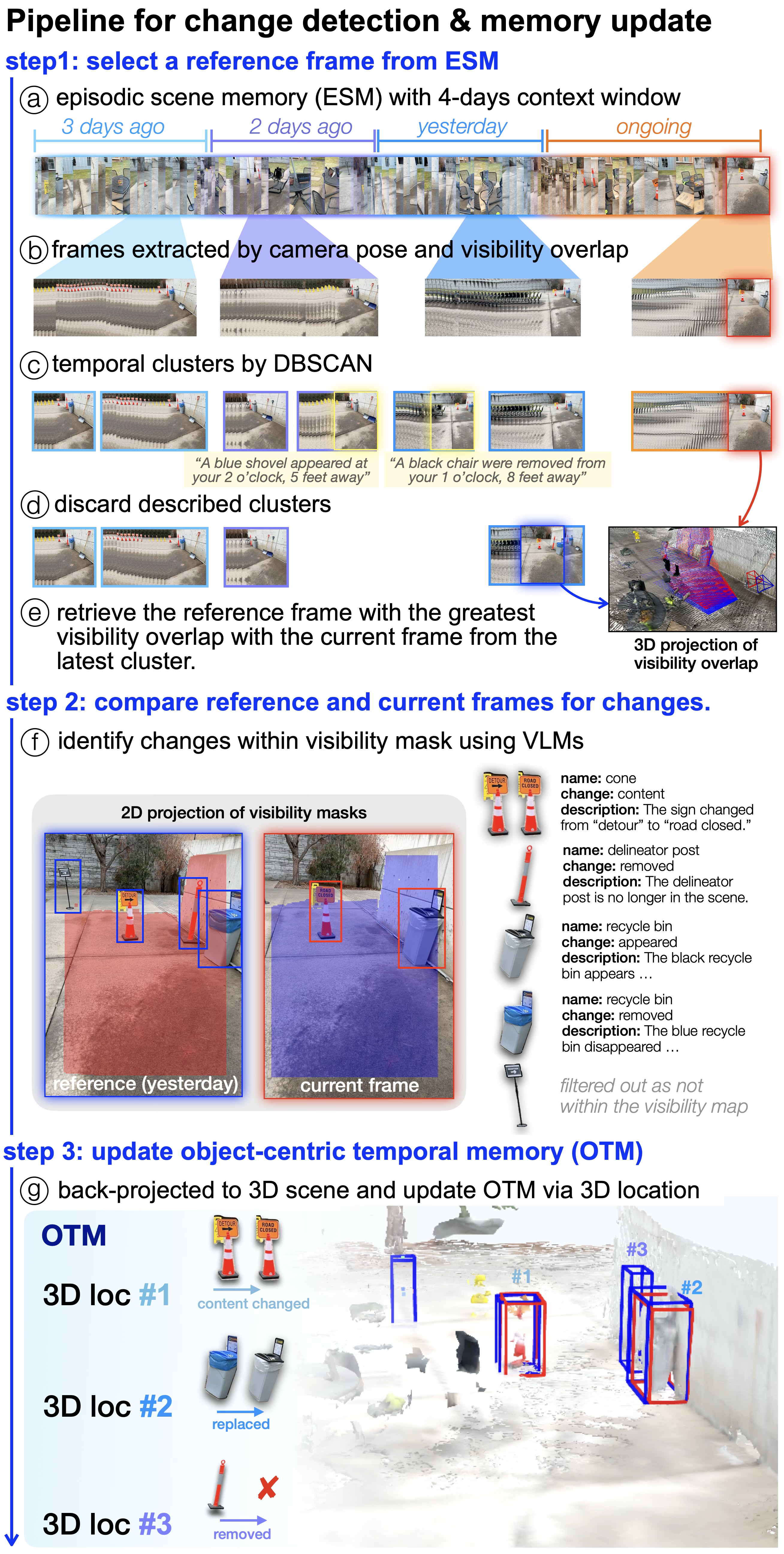}
\vspace{-1.8pc}
\caption{Change detection and memory update pipeline. 
(a) Given the current 
frame (rightmost, outlined in red), (b) {\name} queries its ESM by 
retrieving past frames with similar camera pose and visibility. 
(c) These candidate frames are temporally clustered using DBSCAN. 
(d) Clusters whose changes have been announced are discarded. 
(e) From the most recent valid cluster, {\name} selects the reference 
frame with the highest visibility overlap with the current frame. 
(f) A VLM compares the reference and current frames to detect changes,
producing detailed metadata, such as object identity, change type, 
descriptions, and 2D/3D bounding boxes. 
(g) Finally, the detected changes are back-projected into the 3D 
space to update the OTM.}
\label{fig:keyframe}
\vspace{-1.2pc}
\Description{Figure 3
The figure, titled ``Pipeline for change detection & memory update,'' illustrates a three-step process labeled step 1, step 2, and step 3, with subcomponents (a) through (g), describing how a system selects reference frames, detects changes, and updates memory.
In step 1, titled select a reference frame from ESM,'' panel (a) shows episodic scene memory (ESM) with 4-days context window,'' visualized as a horizontal timeline of image thumbnails labeled 3 days ago,'' 2 days ago,'' yesterday,'' and ongoing.'' Panel (b) shows frames extracted by camera pose and visibility overlap,'' with several distorted or aligned street-view images grouped by viewpoint similarity. Panel (c), titled temporal clusters by DBSCAN,'' groups these frames into clusters, each outlined in blue or orange. Example captions below clusters include: A blue shovel appeared at your 2 o’clock, 5 feet away'' and A black chair were removed from your 1 o’clock, 8 feet away.'' Panel (d), titled discard described clusters,'' shows that previously processed clusters are removed from consideration. Panel (e) reads: retrieve the reference frame with the greatest visibility overlap with the current frame from the latest cluster,'' and includes a visualization labeled ``3D projection of visibility overlap,'' indicating spatial alignment between frames.
In step 2, titled compare reference and current frames for changes,'' panel (f) states: identify changes within visibility mask using VLMs.'' Two images are shown side by side labeled reference (yesterday)'' and current frame,'' each with colored regions indicating visibility masks (e.g., red and blue shaded areas) and bounding boxes around objects. On the right, a list of detected objects and changes is provided. Examples include: name: cone; change: content; description: The sign changed from ‘detour’ to ‘road closed’,'' name: delineator post; change: removed; description: The delineator post is no longer in the scene,'' name: recycle bin; change: appeared; description: The black recycle bin appears,'' and name: recycle bin; change: removed; description: The blue recycle bin disappeared.'' A note at the bottom states ``filtered out as not within the visibility map,'' indicating that only visible regions are considered.
In step 3, titled update object-centric temporal memory (OTM),'' panel (g) shows a 3D reconstructed scene labeled OTM.'' Several objects are marked with 3D bounding boxes and labels such as 3D loc \#1,'' 3D loc \#2,'' and 3D loc \#3.'' Associated change annotations include content changed'' for location \#1, replaced'' for location \#2 (illustrated by two different recycle bins), and removed'' for location \#3 (indicated by a red X over a delineator post). The caption states: ``back-projected to 3D scene and update OTM via 3D location,'' indicating that detected changes are mapped into a spatial memory representation.
}
\end{figure}

\textbf{Object-centric Temporal Memory (OTM).}
To enable persistent tracking of key objects, we introduce OTM, 
an object-centric memory layer that maintains long-term object 
dynamics beyond ESM’s temporal window.
Specifically, OTM maintains a collection of tracked real-world objects, 
each represented as a chronological sequence of discrete snapshots. 
Each snapshot captures the object's state at a specific moment, 
including its status relative to the previous scene (\textit{e.g.,} appeared, 
disappeared, replaced), description, visual embedding, and a world-anchored 
3D bounding box (Figure~\ref{fig:keyframe}f,g).
A new OTM entry is created when a change is detected for an object
(\textit{e.g.,} it appears, disappears, or is replaced). 
When the changed object shares a similar 3D location with a prior OTM object, a new snapshot is appended to the existing OTM entry. These snapshots do not overwrite previous ones; instead, OTM preserves the full history of each tracked object.
We detail this in the next section.

\vspace{0pc}

\begin{figure*}[t]
\begin{center}
\vspace{-1.pc}
\includegraphics[width=\linewidth]{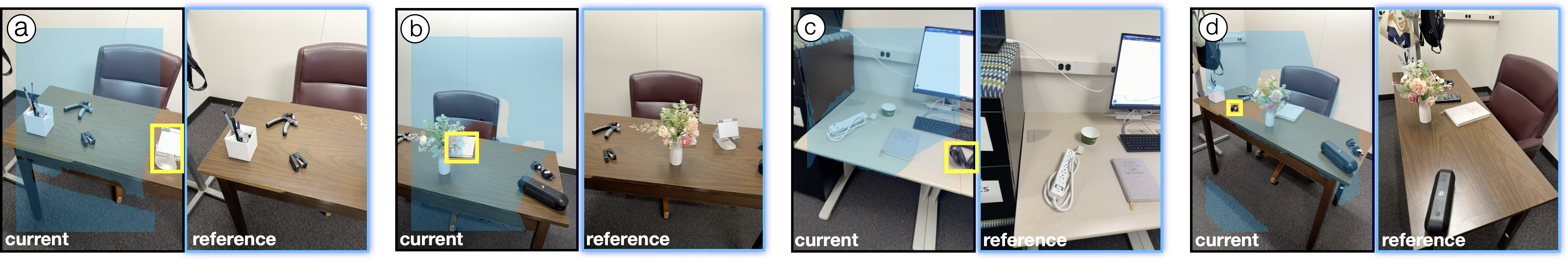}
\vspace{-1.8pc}
\caption{Examples of visibility-mask filtering of potential false positives. The (a) tablet stand, (b) book, (c) headphone, and (d) stapler were not detected as changes because they fell outside the visibility mask.
}
\vspace{-1.pc}
\label{fig:visibilitymask}
\Description{Figure 4
The figure is titled Example use of visibility mask'' and presents four pairs of images, each consisting of a current'' frame and a ``reference'' frame, demonstrating how changes are filtered based on a visibility mask. In each pair, a semi-transparent blue region indicates the visibility mask, and yellow boxes highlight objects of interest.
In the first pair, labeled current'' and reference,'' a desk scene is shown with objects including a tablet stand, a stapler, and other small items. The tablet stand is highlighted with a yellow box. A blue overlay covers part of the desk in the current frame, indicating the visible region. The caption reads: ``The tablet stand was not detected as a change, as it was outside the visibility mask.'' This indicates that although the object may differ between frames, it is ignored because it lies outside the visible region.
In the second pair, another desk scene is shown with a book highlighted in yellow. The blue visibility mask again covers only part of the scene. The caption reads: ``The book was not detected as a change, as it was outside the visibility mask.'' This demonstrates another instance where an object change is excluded due to being outside the visible area.
In the third pair, an office desk setup is shown with a monitor, keyboard, and a set of headphones. The headphones are highlighted with a yellow box in the reference frame. The blue overlay in the current frame does not include the region where the headphones are located. The caption states: ``The headphone was not detected as a change, as it was outside the visibility mask.''
In the fourth pair, a desk scene shows a stapler highlighted in yellow. The blue visibility mask again excludes the region containing the stapler. The caption reads: ``The stapler was not detected as a change, as it was outside the visibility mask.''
}
\end{center}
\end{figure*}

\subsection{Change Detection and Memory Update Module}\label{4_change_detection}
In this module, {\name} identifies changes when a user revisits a location
by comparing the current frame with its ESM and OTM constructed from prior visits. 
The process involves:
\textit{(i)} selecting a reference frame from ESM,
\textit{(ii)} comparing reference and current frames to detect changes,
and \textit{(iii)} updating OTM.
We develop algorithms and integrate lightweight VLMs to balance 
accuracy and latency for live interactions (Section~\ref{4_implementation}).

\textbf{Step 1: Selecting a reference frame from ESM.}
The key challenge is to develop a scalable pipeline to efficiently identify 
the appropriate reference frame from the ESM as it grows over time.
To address this, {\name} utilizes a hierarchical frame retrieval technique.
First, the current camera pose is used as a query to filter and retrieve 
the \textit{ESM data frames} with similar poses, where the differences lie 
within translation ($d_{thres}$) and rotation ($\theta_{thres}$) thresholds. 
Each candidate \textit{ESM data frame} that passes the pose filter is then evaluated against the current camera pose using a bidirectional visibility score, $S_{\textit{overlap}}$.
Specifically, let $P_r$ denote the set of valid 3D points back-projected from 
a reference \textit{ESM data frame} $r$ using its RGB-D data and camera intrinsics, where each 
$\mathbf{p}\in P_r$ represents a 3D point in the reference camera coordinate frame. 
Let $\pi_c(\cdot)$ denote the projection function that maps a 3D point from the 
reference camera coordinate to the current camera frame $c$. 
The resulting projection onto the current frame is referred to as 
the \textit{visibility mask} (Figure~\ref{fig:keyframe}f, right).
We define the coverage ratio of \textit{the visibility mask} over 
the current frame as:
\begin{equation}
o_{r\rightarrow c} = \frac{1}{|P_r|} \sum_{\mathbf{p}\in P_r} 
\mathbf{1}\!\left[\pi_c(\mathbf{p}) \in \Omega_c\right]
\end{equation}
where $\Omega_c$ denotes the valid image domain of the current frame. 
Similarly, $o_{c\rightarrow r}$ is computed in the reverse direction. 
The final bidirectional visibility score is defined as the harmonic mean 
of the two directional coverage ratios:
\begin{equation}
S_{\text{overlap}} = \frac{2\,o_{r\rightarrow c}\,o_{c\rightarrow r}}{o_{r
\rightarrow c}+o_{c\rightarrow r}}
\end{equation}

Second, candidates that satisfy the overlap thresholds are grouped into 
temporal clusters via DBSCAN~\cite{dbscan} based on their timestamps 
(Figure~\ref{fig:keyframe}c). 
Specifically, clustering is performed with a temporal radius of $\epsilon$ 
seconds and a minimum cluster size of $N_{c}$ frames. 
{\name} then scans the clusters in chronological order, discards those whose 
changes have already been announced, and selects the reference frame with the highest visibility score $S_{\textit{overlap}}$ from the latest valid cluster.
In our setup, we set $d_{thres}=1.5$ meters, $\theta_{thres}=40^{\circ}$, 
$\epsilon = 10$, and $N_c = 2$.

\textbf{Step 2: Comparing reference and current frames to detect changes.} Once a 
reference frame is selected, {\name} identifies scene changes by prompting a VLM 
with both the reference and current frames to highlight changed objects using
bounding boxes (Prompt \#\ref{prompt:diff-system}).
It returns the type of change for each detected object, including appeared, 
removed, or content changed, along with a textual description and a confidence score. 
{\name} then rejects bounding boxes whose overlap with their corresponding visibility 
masks (described in step 1) falls below a threshold $X$ (Figure~\ref{fig:visibilitymask}).
The remaining 2D bounding boxes are further filtered by confidence and size, then 
passed to a segmentation model to extract object masks and generate visual embeddings. 
These masks are subsequently projected into 3D space to estimate the objects' 
3D locations. 
Our current setup processes at 1 FPS, and we empirically set the temporal clustering parameters to $X=0.45$.

\textbf{Step 3: Updating OTM.}
Each detected change, along with its 2D mask, 3D bounding box, and the associated 
change type and description, is then updated in OTM. We determine if a detected 
object is new or corresponds to an existing OTM entry by computing the 3D Intersection 
over Union (IoU) between the detected 3D bounding box $B_{new}$ and a prior 3D 
bounding box $B_{prior}$:
\begin{equation}
\text{IoU}(B_{new}, B_{prior}) = \frac{V_{int}}{V_{new} + V_{prior} - V_{int}}
\end{equation}
where $V_{int}$ denotes the intersection volume of the two boxes. {\name} iterates 
over existing objects in OTM and retains candidate prior boxes $B_{prior}$ 
whose IoU with $B_{new}$ exceeds a minimum threshold $\gamma$. For the remaining 
candidates, {\name} computes the cosine similarity between the visual embeddings of 
the new and prior objects. If the similarity exceeds a threshold $Y$, the object is 
treated as the same instance and recorded as a new snapshot in the corresponding OTM 
entry; otherwise, it is stored as a new OTM entry. Identified change snapshots, 
including change type, descriptions, reference, and current images, are then queued in the scheduling module to present (Section~\ref{4_scheduling_layer}). 
We set $\gamma=0.08$ and $Y=0.7$.

\subsection{Live Description Generation Module}
In this module, {\name} generates live visual descriptions to support scene 
understanding. Unlike prior systems that use multiple VLMs and adaptive prompting 
for context-aware descriptions~\cite{WorldScribe, TouchScribe}, {\name} adopts a simplified approach using frame filtering and a single VLM with a fixed prompt 
(Prompt \#\ref{prompt:live-scene}).
To detect new scenes, it computes visual embeddings for incoming frames and 
compares them with the most recently described frame using cosine similarity; 
frames exceeding a threshold ($\tau_{\text{visual}}$) are treated as redundant 
and skipped. 
To further reduce repetition, generated descriptions are also compared via text 
embeddings with prior outputs, and those exceeding a threshold 
($\tau_{\text{text}}$) are discarded. The remaining descriptions are queued 
for presentation (Section~\ref{4_scheduling_layer}). 
We empirically set $\tau_{\text{visual}}=0.85$ and $\tau_{\text{text}}=0.80$ by default.

\subsection{Interactive Q\&A Module}
In this module, {\name} allows users to interrupt ongoing descriptions and issue 
queries. It supports change-related queries over single or multiple objects 
across time, within specified intervals, and their spatial variations.
This is achieved by a Q\&A agent that interprets each query and invokes tools over
ESM, OTM, and spatial data (Prompt \#\ref{prompt:agent-qa}): \textit{(i)} 
\tighttt{ESM retrieval} for current scene queries, returning the 
$N$ most recent frames;
\textit{(ii)} \tighttt{OTM retrieval} for temporal changes, returning recent 
object snapshots with distance and direction; and 
\textit{(iii)} \tighttt{Spatial information retrieval} for spatial queries, 
returning object locations (\textit{e.g., ``11 o'clock, 5 feet away''}).
These tools are composed based on query intent. For example, \textit{``Where is 
the cone?''} uses \tighttt{ESM retrieval} and \tighttt{Spatial information retrieval}, 
while \textit{``What changed since my last visit?''} uses \tighttt{OTM retrieval}.

\subsection{Description Scheduling \& Delivery Module}\label{4_scheduling_layer}
In this module, {\name} coordinates change snapshots, live descriptions, and interactive Q\&A. Visual queries are prioritized, after which live and change descriptions resume. {\name} maintains a buffer ($N=3$) of recent live descriptions and change snapshots.
Live descriptions that persist beyond a time threshold (\textit{e.g.,} 6 seconds) are
considered outdated and discarded, while change snapshots (\textit{e.g.,} \textit{appeared}, 
\textit{removed}, \textit{content changed}), along with their images and 
3D locations, are retained. 
Using this aggregated context in a single prompt (\#\ref{prompt:scene-paraphrase}), 
{\name} employs a VLM to infer higher-level events, such as object replacement 
(via overlapping 3D positions) or movement (via disappearance and reappearance 
across locations).

\subsection{Mobile User Interface}\label{4_mobile_interface}
{\name} provides a mobile app interface (Figure~\ref{fig:pipeline}a). 
The home screen displays visits as cards, each with a thumbnail and an AI-generated title.
When returning to a location, users can select the corresponding card to open 
a live camera view for relocalization and resume {\name}'s memory for that place. 
For new locations, they can create a visit using the ``Add'' button. Users can also 
rename or delete visits.
While the current interface requires users to manually recall and select past visits, 
future iterations could incorporate GPS or indoor localization data to automatically 
cluster visits or create a new visit upon arrival at a new location.

\subsection{Implementation Details}\label{4_implementation}
To enable {\name}’s spatial and temporal memory capabilities, we use a smartphone 
as the primary device to access depth information, which is not available on the current smartglasses that typically rely on monocular cameras. We also adopt a low frame rate (i.e., 1 FPS) to support long-term use and scalability for memory storage, retrieval, and live interaction, in contrast to the higher frame rates used in prior systems~\cite{WorldScribe, TouchScribe}.
Our {\name} mobile app is deployed on an iPhone 17 Pro and relies on camera poses 
and RGB-D images are continuously captured via Apple’s ARKit framework. 
These data streams are transmitted to the backend at one FPS over a TCP connection. 
{\name} leverages ARKit’s built-in geometric relocalization to align the current camera coordinates with those from prior visits to the same location.
Each location is assigned a unique identifier by the mobile application. 
Upon successful relocalization, the backend uses this identifier to load the corresponding ESM and OTM.
The backend is hosted on a MacBook M4 Max for our user study, while the technical evaluation is conducted on a desktop with an AMD Ryzen Threadripper PRO 7965WX CPU and an NVIDIA RTX 6000 Ada GPU.

We adopt lightweight VLMs for live interactions to balance latency and accuracy. Specifically, live descriptions are generated using \tighttt{Gemini 3.1 Flash-Lite Preview}, while change detection, Q\&A, and buffered description summarization are handled by \tighttt{Gemini 3 Flash Preview}. FastSAM~\cite{zhao2023fast} is used for 2D mask segmentation. Visual embeddings are extracted using a pre-trained DINOv3 model~\cite{simeoni2025cijo}, and text embeddings are generated with \tighttt{all-MiniLM-L12-v2}.
Finally, we limited the ESM context window to the previous visit for both user and technical evaluations.

\section{Technical Evaluation}
We evaluated whether {\name} \textit{(i)} preserves longitudinal memory across visits, \textit{(ii)} scales efficiently with growing data, and \textit{(iii)} improves real-time change awareness in 3D environments. We focused on the live description pipeline---change detection, memory updates, and announcements---and excluded the auxiliary Q\&A module because participants’ questions varied and were not always change-related. Evaluating the Q\&A module remains future work.

\subsection{Datasets}
\begin{figure}[h]
\begin{center}
\vspace{-0.5pc}
\includegraphics[width=\linewidth]{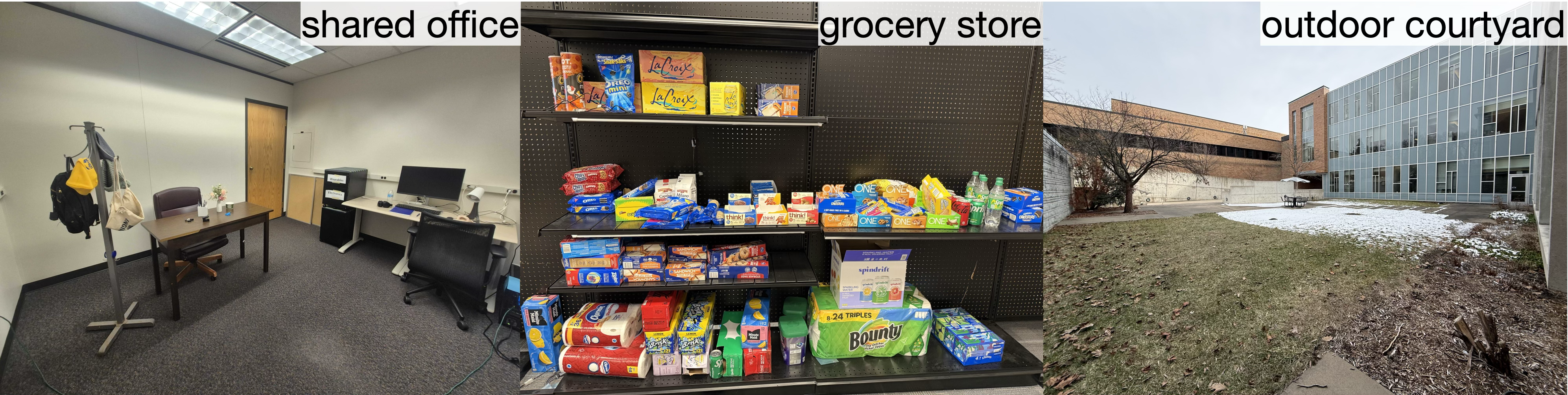}
\vspace{-1.8pc}
\caption{Our study environments, including a shared office, a simulated grocery store, and an outdoor courtyard.}
\vspace{-1.2pc}
\label{fig:studyenv}
\Description{Figure 5
The figure shows three side-by-side photographs labeled shared office,'' grocery store,'' and ``outdoor courtyard,'' each depicting a different environment.
In the left image labeled ``shared office,'' a small office room is shown with beige walls and a drop ceiling with fluorescent lighting panels. A wooden desk is positioned near the center-left, with a chair tucked underneath it. On the desk are small items including a plant and office supplies. To the right, a second desk holds a computer monitor, keyboard, and other office equipment, with a black office chair in front of it. A coat rack stands on the left side of the room, holding several bags and a jacket. A door with a wooden finish is visible in the back wall, and the floor is covered with dark carpet.
In the middle image labeled ``grocery store,'' a set of black metal shelves is filled with various packaged food items. The shelves contain colorful boxes, bags, and containers, including snacks, cookies, and other grocery products. Items are arranged in multiple rows across three visible shelf levels. The packaging includes bright colors such as yellow, blue, red, and orange, with visible brand-style text and graphics, though specific product names are not clearly readable. The scene is tightly framed, focusing on the stocked shelves.
In the right image labeled ``outdoor courtyard,'' an open outdoor space is shown between modern buildings. The ground consists of grass with some patches of snow scattered across it. On the right side, a multi-story building with large windows and a glass facade is visible, while on the left side there is a lower building with a more solid exterior. A leafless tree stands near the center-left, indicating a colder season. The sky appears overcast, and the overall scene has muted lighting consistent with winter conditions.
}
\end{center}
\end{figure}

To simulate extended use, we constructed a dataset of repeated visits to the same environments over time. The dataset includes three environments (Figure~\ref{fig:studyenv}): \textit{(i)} a shared office with varying objects, \textit{(ii)} a grocery store with densely arranged items across shelves, and \textit{(iii)} an outdoor courtyard featuring signs and barriers. Each scenario contains 11 recorded visits (around 2--4 minutes per visit), including a video, RGB-D frames, camera intrinsics, and poses, capturing observations at different time points. 
We systematically introduced object changes based on the spatial layout of the shared office, 104 for the grocery, and 83 for the outdoor scenario. 
In total, the dataset contains 291 annotated changes, labeled by three researchers with 3D world locations, object identities, and change categories. 

\begin{figure*}[t]
\centering
\vspace{-1.2pc}
\includegraphics[width=\linewidth]{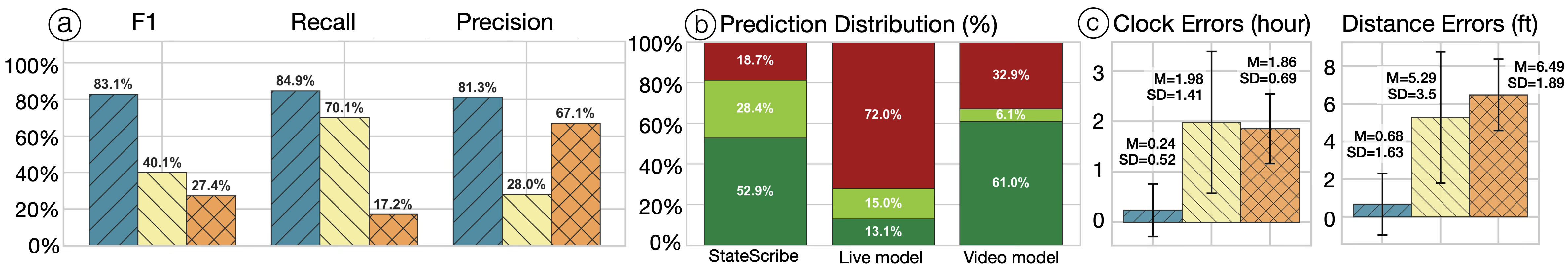}
\noindent\textbf{(i) {\name}'s overall performance on our collected dataset.}
\includegraphics[width=\linewidth]{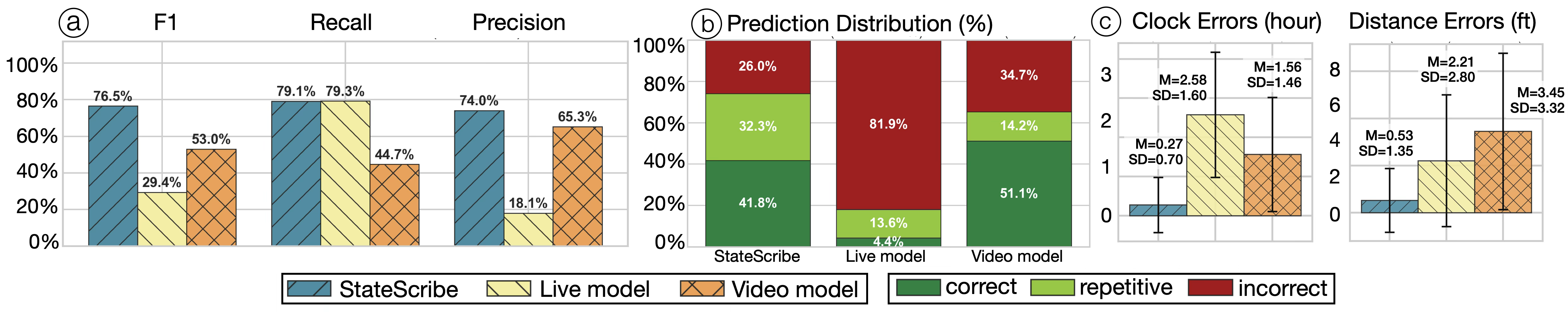}
\noindent\textbf{(ii) {\name}'s overall performance in the user study. }

\vspace{-1pc}
\caption{{\name}'s overall performance across two evaluation settings:
\textit{(i)} technical evaluation with our collected datasets and \textit{(ii)} user study with data collected by nine BLV participants.
Both use the same evaluation metrics:
(a) performance metrics including \textit{F1 score}, \textit{recall}, and \textit{precision} across models;
(b) \textit{prediction distribution} for {\name}, Live model, and Offline video model,
showing the number of correct, repetitive, and incorrect outputs;
and (c) error analysis showing mean and standard deviation for \textit{Clock} and
\textit{Distance Errors}.}
\label{fig:overall_results_combined}
\Description{Figure 6
The above figure contains three panels labeled (a), (b), (c), presenting evaluation results comparing three systems: StateScribe, Live model, and Video model. The metrics include F1 score, recall, precision, prediction distribution, and error measurements.
In panel (a), three grouped bar charts are shown for F1,'' Recall,'' and ``Precision,'' each with percentages on the vertical axis from 0\% to 100\%. For F1, the values are StateScribe (83.1\%), Live model (40.1\%), and Video model (27.4\%). For Recall, the values are StateScribe (84.9\%), Live model (70.1\%), and Video model (17.2\%). For Precision, the values are StateScribe (81.3\%), Live model (28.0\%), and Video model (67.1\%). Across all three metrics, StateScribe consistently achieves the highest scores.
In panel (b), titled ``Prediction Distribution (\%),'' three stacked bars represent the proportion of predictions categorized as correct, repetitive, and incorrect. For StateScribe, the distribution is 52.9\% correct (dark green), 28.4\% repetitive (light green), and 18.7\% incorrect (red). For the Live model, the distribution is 13.1\% correct, 15.0\% repetitive, and 72.0\% incorrect. For the Video model, the distribution is 61.0\% correct, 6.1\% repetitive, and 32.9\% incorrect. These bars show that StateScribe and the Video model have higher proportions of correct predictions compared to the Live model, which has a large proportion of incorrect outputs.
In panel (c), titled ``Clock Errors (hour),'' a bar chart shows the mean (M) and standard deviation (SD) for each system. StateScribe has M = 0.24 and SD = 0.52, the Live model has M = 1.98 and SD = 1.41, and the Video model has M = 1.86 and SD = 0.69. StateScribe exhibits the lowest clock error.
Another panel in (c), titled ``Distance Errors (ft),'' another bar chart shows mean and standard deviation values. StateScribe has M = 0.68 and SD = 1.63, the Live model has M = 5.29 and SD = 3.5, and the Video model has M = 6.49 and SD = 1.89. Again, StateScribe shows substantially lower error compared to the other models.
The below figure contains three panels labeled (a), (b), and (c), comparing performance metrics across three systems: StateScribe, Live model, and Video model. The metrics include F1 score, recall, precision, prediction distribution, and error measurements.
In panel (a), three grouped bar charts are shown for F1,'' Recall,'' and ``Precision,'' with percentages on the vertical axis from 0\% to 100\%. For F1, the values are StateScribe (76.5\%), Live model (29.4\%), and Video model (53.0\%). For Recall, the values are StateScribe (79.1\%), Live model (79.3\%), and Video model (44.7\%). For Precision, the values are StateScribe (74.0\%), Live model (18.1\%), and Video model (65.3\%). StateScribe has the highest F1 and precision, while recall is similar between StateScribe and the Live model.
In panel (b), titled ``Prediction Distribution (\%),'' three stacked bars show the proportions of predictions categorized as correct, repetitive, and incorrect. For StateScribe, the distribution is 52.9\% correct, 28.4\% repetitive, and 18.7\% incorrect. For the Live model, it is 13.1\% correct, 15.0\% repetitive, and 72.0\% incorrect. For the Video model, it is 61.0\% correct, 6.1\% repetitive, and 32.9\% incorrect. This indicates that the Live model produces the highest proportion of incorrect predictions, while StateScribe and the Video model perform better.
In panel (c), two bar charts are shown. The first, Clock Errors (hour),'' reports mean (M) and standard deviation (SD): StateScribe (M = 0.27, SD = 0.70), Live model (M = 2.58, SD = 1.60), and Video model (M = 1.56, SD = 1.46). The second, Distance Errors (ft),'' shows StateScribe (M = 0.53, SD = 1.35), Live model (M = 2.21, SD = 2.80), and Video model (M = 3.45, SD = 3.32). StateScribe has the lowest errors in both measures.
}
\vspace{-1.2pc}
\end{figure*}

\subsection{Baseline Conditions and Evaluation Metrics} 
To evaluate the effectiveness of {\name}'s design, we compared it against two VLM-only baselines: a \textit{Live model} and an \textit{Offline video model}.
For both baselines, we concatenated the recorded videos from multiple visits in the same scenario and sampled them at 1 FPS.
In the \textit{Live model},  we prompted the VLM (i.e., \tighttt{gemini-3-flash-preview}) at 1 FPS to approximate real-time interaction similar to {\name} (Prompt \#\ref{prompt:online-baseline}).
This baseline examines whether {\name} improves over a direct VLM-only real-time pipeline.
In the \textit{Offline video model} condition, we provided the concatenated 1 FPS video to \tighttt{gemini-3.1-pro} as a single input for detecting changes within or across visits (Prompt \#\ref{prompt:offline-baseline}).
This baseline evaluates {\name} against an offline setting that processes all observations jointly rather than sequentially.
For both baselines, we used the same Gemini model family as in {\name} to isolate the effect of system design rather than differences across model families.

We used \tighttt{Gemini 3.1 Pro} as an LLM judge to compare predicted objects, locations, change types, and descriptions (e.g., \textit{``The white Milano cookie box has been removed from the shelf.''}) with ground truth. The judge assessed whether each description captured the ground-truth event, rather than its standalone quality. Nevertheless, this design may not fully eliminate same-family bias, although a preliminary cross-family validation using GPT-5.5 (rather than a Gemini-family model) as the judge yielded similar trends. Future work could further validate these findings through additional human ratings.
We reported \textit{precision}, \textit{recall}, and \textit{$F_1$-score}, and analyzed output counts for \textit{repetitiveness} and \textit{coverage} (Section~\ref{5_overall}). Spatial accuracy, including \textit{clock} and \textit{distance} errors, was computed directly from annotated 3D boxes, camera poses, and ground truth without the LLM judge. We also repeated this evaluation on data from BLV participants (Section~\ref{5_blind_data}). 
To assess scalability, we simulated 110 visits by repeating an 11-visit sequence 10 times and measured \textit{storage} and \textit{latency} (Section~\ref{7_longterm}).

\subsection{Results}
\begin{figure}[h]
\centering
\vspace{-0.5pc}
\includegraphics[width=\linewidth]{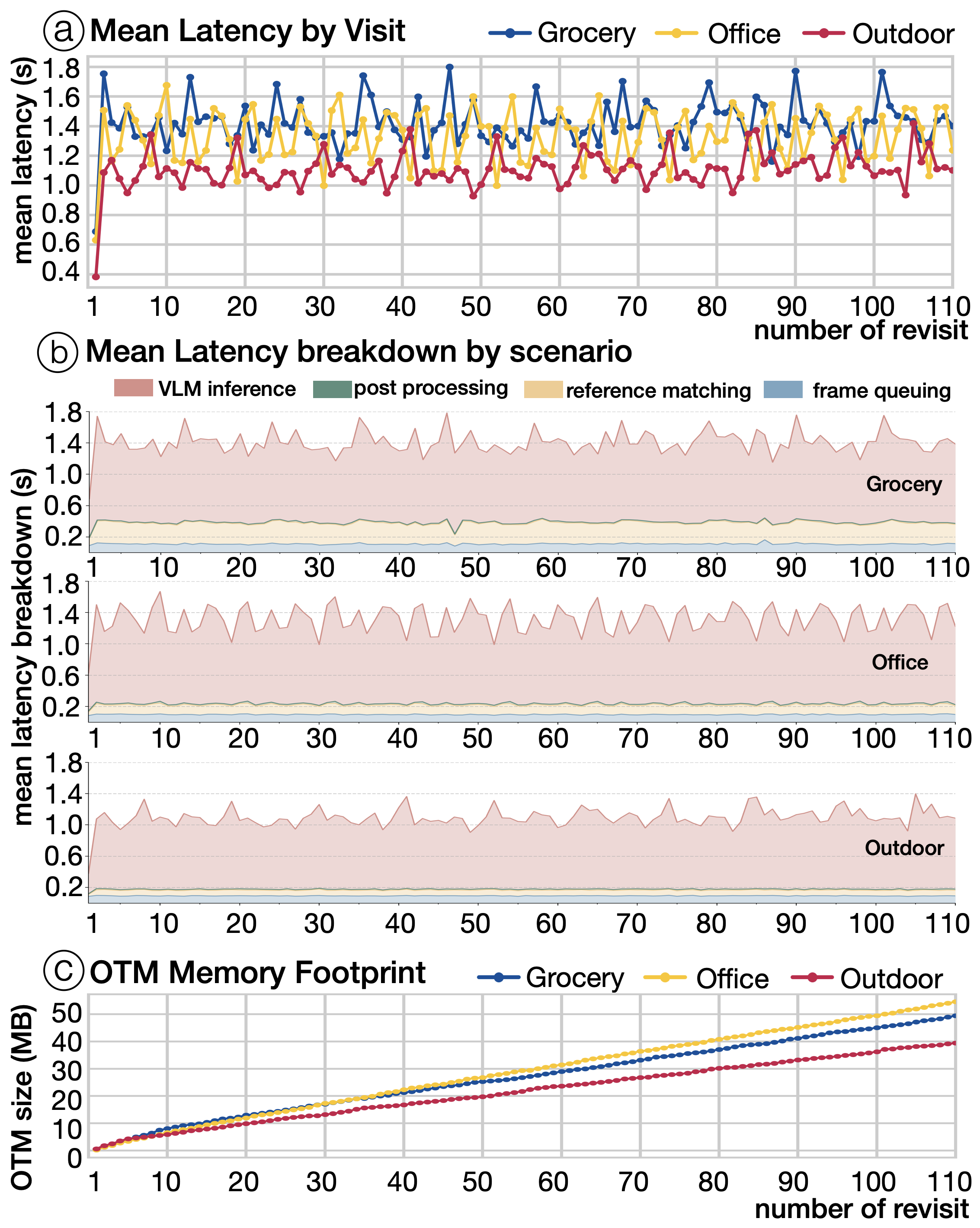}
\vspace{-2pc}
\caption{{\name}’s (a) mean latency, (b) their breakdown across system components, and 
(c) memory footprint over 110 visits in three scenarios: \textit{Grocery}, \textit{Office}, 
and \textit{Outdoor}.}
\label{fig:longterm}
\vspace{-1pc}
\Description{Figure 7
The figure contains three panels labeled (a), (b), and (c), showing system latency and memory usage across different scenarios labeled Grocery, Office, and Outdoor.
In panel (a), titled ``Mean Latency by Visit,'' a line chart plots mean latency in seconds on the vertical axis (ranging approximately from 0.4 to 1.8 seconds) against the number of revisits on the horizontal axis (from 1 to 110). Three lines are shown: Grocery (blue), Office (yellow), and Outdoor (red). The Grocery and Office lines fluctuate mostly between about 1.2 and 1.7 seconds, while the Outdoor line is generally lower, fluctuating roughly between 1.0 and 1.3 seconds. All three lines show variability across visits but no strong upward or downward trend, indicating relatively stable latency over time.
In panel (b), titled Mean Latency breakdown by scenario,'' three stacked area plots are shown vertically, one for each scenario: Grocery, Office, and Outdoor. The vertical axis represents mean latency breakdown in seconds (up to about 1.8 seconds), and the horizontal axis again represents the number of revisits (1 to 110). Each plot is composed of four colored layers representing different components: VLM inference'' (largest portion, in a reddish color), post processing,'' reference matching,'' and ``frame queuing'' (smallest portion). Across all scenarios, VLM inference dominates the total latency, while the other components contribute relatively small and stable portions. The overall shape remains consistent across visits, with slight fluctuations.
In panel (c), titled ``OTM Memory Footprint,'' a line chart shows memory usage in megabytes (MB) on the vertical axis (from 0 to about 55 MB) versus number of revisits on the horizontal axis (1 to 110). Three lines are plotted: Grocery (blue), Office (yellow), and Outdoor (red). All three lines increase steadily over time, indicating growing memory usage as revisits accumulate. The Office scenario shows the highest memory usage, reaching around 50 MB, followed by Grocery at approximately 45 MB, and Outdoor at around 40 MB by the final revisit.
}
\end{figure}

We first report the overall performance across models on our collected dataset
(Figure~\ref{fig:overall_results_combined}.i) and on data captured by BLV
participants in our user study (Figure~\ref{fig:overall_results_combined}.ii).
We then evaluate the performance of {\name} under extended use (Figure~\ref{fig:longterm}).

\subsubsection{Overall Performance}\label{5_overall}
\begin{figure*}[t]
\begin{center}
\vspace{-1pc}
\includegraphics[width=\linewidth]{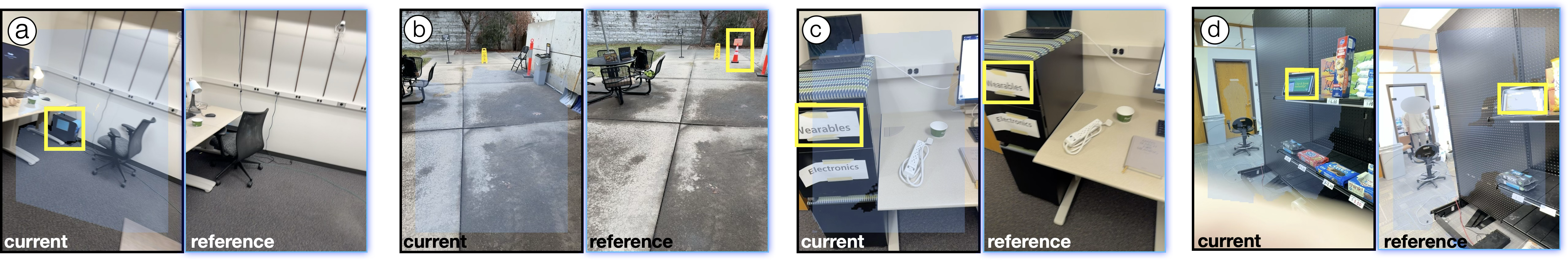}
\vspace{-1.8pc}
\caption{StateScribe errors. (a) Objects that are present but occluded before may be incorrectly classified as ``appeared.'' (b) The visibility mask may be truncated for regions outside the LiDAR coverage. (c) Truncated text may be misinterpreted as an attribute change, such as from ``Wearables'' to ``Nearables.'' (d) Light reflections may also be mistaken for attribute changes.}
\vspace{-1.2pc}
\label{fig:errors}
\Description{Figure 8
The figure presents StateScribe errors and presents four cases labeled (a), (b), (c), and (d), each showing a pair of images labeled current'' and ``reference,'' along with explanations of incorrect change detections. Yellow boxes highlight regions where false positives occur.
In panel (a), two office scenes are shown. In the reference'' image, a chair and desk are clearly visible, while in the current'' image, part of the scene is occluded, and a small object near the desk is highlighted with a yellow box. The caption reads: ``objects occluded but present in current view will be mistaken as ‘appeared’.'' This indicates that occlusion causes the system to incorrectly detect a new object when it was already present but previously hidden.
In panel (b), two outdoor pavement scenes are shown. The reference'' image includes objects such as cones and barriers, with a highlighted region in yellow. The current'' image shows a similar area but with a semi-transparent overlay indicating limited coverage. The caption states: ``the visibility mask may be cut off for the parts outside LiDAR coverage.'' This demonstrates that incomplete visibility leads to incorrect detection of changes.
In panel (c), two desk scenes are shown with labels on drawers or boxes. The highlighted region focuses on text labels, where the wording appears slightly different between frames (e.g., Wearables'' versus Nearables''). The caption reads: ``text cutoffs could be mistaken as an attribute change, from ‘Wearables’ to ‘Nearables’.'' This shows that partial text visibility can lead to misinterpretation of changes.
In panel (d), two grocery shelf scenes are shown. The highlighted region includes a reflective surface or brightly lit area. The caption reads: ``light reflection could be mistaken as an attribute change.'' This indicates that lighting variations or reflections may be incorrectly interpreted as meaningful changes.
}
\end{center}
\end{figure*}

We reported the overall performance across all 291 ground-truth instances 
collected from three scenarios, each with 11 visits.
{\name} outperformed both baselines in change detection 
(Figure~\ref{fig:overall_results_combined}.i-a). 
It achieved an $F_1$-score of \textbf{83.1\%}, substantially higher than the 
\textit{Live model} (40.1\%) and the \textit{Office video model} (27.4\%). 
This performance was driven by both high \textit{recall} (84.9\%) and \textit{precision} 
(81.3\%), indicating that {\name} captures most changes while avoiding excessive false 
positives.
In contrast, the \textit{Live model} achieved moderate \textit{recall} (70.1\%) 
but low \textit{precision} (28.0\%), resulting in many incorrect predictions. 
The \textit{Offline video model}, while achieving moderate \textit{precision} (67.1\%), 
exhibited low \textit{recall} (17.2\%), missing detecting most changes that lead to 
poor overall performance.
Most false positives in both models occurred when objects appeared near the image edge
without prior visibility (Figure~\ref{fig:modelerror}). Likewise, objects misclassified 
as \textit{removed} had merely left the camera’s view while still existing 
(Figure~\ref{fig:modelerror}), whereas {\name} mitigated this using visibility 
masks (Figure~\ref{fig:visibilitymask}).
Errors in {\name} mainly arose for distant objects beyond LiDAR coverage, text cutoffs or light reflections (Figure~\ref{fig:errors}).

When examining the \textit{Prediction Distribution} across models, including counts of 
correct, incorrect, and repetitive predictions, we observed a wide variance in output 
behavior (Figure~\ref{fig:overall_results_combined}.i-b).
{\name} produced 444 predictions, including 81.13\% correct (N=361, including 126 
repetitive), and 18.7\% incorrect outputs (N=82).
In contrast, the \textit{Live model} produced a large number of predictions (N=1558), 
despite only 291 true changes in the dataset. As a result, the majority of its outputs
were incorrect (72.0\%, N=1121), while 28.1\% were correct (N=437, including 233 repetitive), 
reflecting both a high error rate and considerable redundancy.
Furthermore, the \textit{Offline video model} generated far fewer predictions (N=82), 
comprising 61\% correct (N=50, including 5 repetitive), and 32.9\% incorrect outputs (N=27). 
This aligns with its previously observed high \textit{precision} and low \textit{recall}, 
suggesting conservative behavior or limited ability to capture all changes.

In terms of spatial accuracy (Figure~\ref{fig:overall_results_combined}.i-c), {\name} outperformed 
both models, achieving lower clock-direction errors (in hour units; M=0.24, SD=0.52) than the \textit{Live model} (M=1.98, SD=1.41) and the \textit{Offline video model} (M=1.86, SD=0.69). A similar trend was observed for distance estimation: {\name} yielded smaller errors (in feet; M=0.68, SD=1.63) than the \textit{Live model} (M=5.29, SD=3.50) and the \textit{Offline video model} (M=6.49, SD=1.89).

Overall, {\name} achieved a better balance between coverage and accuracy, identifying more 
changes than both models while having substantially fewer errors and repetitive outputs than 
the \textit{Live model}. This trend also extended to spatial accuracy, with {\name} showing 
lower clock-direction and distance estimation errors.

\subsubsection{StateScribe's performance in user study}\label{5_blind_data}
Next, we used the same evaluation metrics on the data collected from our user study (see 
Section~\ref{6_procedure} for data details), where camera frames from BLV participants 
may differ due to camera aiming. 
Overall, {\name} showed similar trends (Figure~\ref{fig:overall_results_combined}.ii), 
outperforming both baselines across 269 tasks. 
It achieved an $F_1$-score of \textbf{76.5\%}, higher than the \textit{Live model} (29.4\%)
and the \textit{Offline video model} (53.0\%). This performance is driven by both high 
\textit{recall} (79.1\%) and \textit{precision} (74.0\%), indicating effective change
detection with limited false positives.
In comparison, the \textit{Live model} achieved similar \textit{recall} (79.3\%) but 
very low \textit{precision} (18.1\%), leading to many false positives. 
The \textit{Offline video model} showed moderate \textit{precision} (65.3\%) but 
low \textit{recall} (44.7\%), missing many changes.
Similar trends were observed in spatial accuracy. {\name} achieved lower clock-direction 
error (in hours; M=0.27, SD=0.70) than the \textit{Live model} (M=2.58, SD=1.60) and the 
\textit{Offline video model} (M=1.56, SD=1.46). It produced smaller distance errors 
(in feet; M=0.53, SD=1.35) compared to the \textit{Live model} (M=2.21, SD=2.80) 
and the \textit{Offline video model} (M=3.45, SD=3.32).
Post-hoc analysis suggests that the performance difference between datasets collected by sighted and BLV participants (Figure~\ref{fig:overall_results_combined}i-a \& ii-a) stems from movement patterns. Sighted participants tended to move faster, whereas BLV participants lingered longer in each scene despite less stable camera handling, giving the video model more opportunities to detect changes and yielding higher recall and F1 scores. Nevertheless, {\name} outperformed the baseline models on both datasets.

\subsubsection{{\name}'s latency and memory footprint over extended use}\label{7_longterm}
To model long-term use and measure latency and memory growth, we repeated 11 visits 10 times (110 visits per scenario), producing 17,350 frames for Office, 17,460 frames for Grocery, and 18,360 frames for Outdoor. Each scenario corresponds to about 5 hours at 1 FPS. This setup is valid because {\name}’s context window included only the prior visit (Section~\ref{4_implementation}), so every visit triggered the full change-detection and memory-update pipeline.

In terms of \textit{latency}, it remains low across all scenarios, including Grocery (Median=0.37s, Mean=1.42s), Office (Median=0.21s, Mean=1.32s), and Outdoor (Median=0.14s, Mean=1.11s), which reflected long-tailed distributions with occasional outliers. 
This indicated that most interactions were processed quickly, with only rare spikes.
To further examine latency contributions, we decomposed processing time into system components. 
Across all scenarios, \textit{VLM inference} dominated latency (Grocery: M=1.02s; Office: M=1.08s; Outdoor: M=0.91s). In contrast, \textit{frame queuing} for processing (M=0.11s, 0.10s, 0.09s), \textit{reference matching} for retrieving past frame data (M=0.27s, 0.12s, 0.07s), and \textit{post-processing} for updating OTM (M=0.01s across all scenarios) contributed relatively minor overhead. 
Notably, reference matching was higher in Grocery (M=0.27s) than in Office (M=0.12s) and Outdoor (M=0.07s), likely due to the denser scenes increasing matching complexity.
Overall, these results suggest that {\name} maintains stable performance over extended use, with latency primarily bounded by VLM inference rather than system-level overhead.

Over extended use for 110 visits, {\name} maintained a bounded memory footprint that scales proportionally with the number of visits. The final OTM size across 110 visits was 49.4 MB (Grocery), 54.6 MB (Office), and 39.4 MB (Outdoor). The Office scenario exhibited the largest memory due to more change snapshots (N=1,517) of tracked objects (N=204), while Grocery showed moderate growth with moderate snapshots (N=1,107) of tracked objects (N=192), and Outdoor remained smaller due to fewer snapshots (N=999) of tracked objects (N=174).
Overall, latency did not increase with OTM size, as our memory architecture reads only prior-visit data from ESM, discards outdated information, and stores only essential object changes in OTM. The current OTM scale (110 visits) also remains computationally manageable during retrieval.

\section{User Study}
Next, we conducted a user study with 9 BLV participants to examine:
\textit{(i)} How effectively does {\name} support users in perceiving and understanding changes in different environments? 
\textit{(ii)} How do participants perceive the user experience of {\name}?

Participants, 6 Male and 3 Female, were recruited through prior contacts and snowball sampling (Table~\ref{tab:demographic-userstudy}).
They aged 25--74 (mean: 49.7); 6 identified as blind and 3 as having low vision.
Most had prior experience with RSA services (e.g., Be My Eyes~\cite{bemyeyes}, Aira~\cite{aira}) and AI tools (e.g., Seeing AI~\cite{seeingai}, Be My AI~\cite{bemyai}).

\subsection{Procedure, Tasks and Analysis}\label{6_procedure}
The study consisted of three sessions:
\textit{(i)} an \textit{onboarding session}, where participants learned to use {\name};
\textit{(ii)} a \textit{main task session}, where participants were asked to identify 
changes in three different scenarios, including a shared office, a grocery store, 
and an outdoor courtyard; and
\textit{(iii)} an \textit{interview session}, where participants reflected on their 
experiences and responded to Likert-scale questions.  

For each scenario in the \textit{main task session}, participants first explored the 
environment and then revisited it. During each visit, researchers introduced 
\textit{within-visit changes} by moving objects in real time to simulate common dynamics
in shared or public spaces (e.g., placing or removing items; Table~\ref{tab:visit_setup}). 
Examples included placing a bottle on a shared table, taking items in the grocery store, 
or removing outdoor obstacles by workers (Table~\ref{tab:visit_setup}).
Between visits, the researcher introduced \textit{cross-visit changes} to simulate 
longer-term rearrangements (\textit{e.g.,} days or weeks). 
These included moving chairs in the shared office, price changes in the grocery store, 
and removing signage in the outdoor walkway.
These changes approximated everyday dynamics within the time constraint of a two-hour study.
In total, the shared office and grocery store scenarios each included 13 changes 
(8 within-visit, 5 cross-visit), while the outdoor scenario included 12 changes
(8 within-visit, 4 cross-visit).

Participants independently explored the indoor environments, including a shared office and a grocery store. In the outdoor scenario, a researcher accompanied participants for safety but did not interfere with their exploration.
In total, we collected 18 visit records for the shared office (117 tasks = 9 participants
× 13 changes), 16 for the grocery store (104 tasks = 8 participants × 13 changes; 
P9 excluded due to corrupted data), and 8 for the outdoor courtyard 
(48 tasks = 4 participants × 12 changes), as only four participants completed the outdoor condition due to varying weather conditions. In total, this yielded 269 tasks.
During each visit, participants thought aloud as they identified changes. We evaluated task completion, Likert-scale ratings, and qualitative feedback, marking tasks as successful when changes were accurately described.

\subsection{Results}\label{6_results}

\subsubsection{\textbf{How effectively does {\name} support users in perceiving 
and understanding changes in different environments?}}\label{6_RQ1}
Overall, participants successfully completed most tasks within around 10 minutes, achieving 
completion rates of 82.9\% (shared office), 82.7\% (grocery), and 89.6\% (outdoor).
In the shared office scenario, issues were mainly due to camera aiming: some regions 
(e.g., parts of desks) were not captured, leading to incomplete memory, and subtle 
view changes, such as slightly pulled-out chairs, were hard to detect (4/9 misses).
In the grocery scenario, removing the cart from the aisle had a lower completion rate (50\%) 
than in other tasks, as participants focused the camera on nearby shelf items, 
and often missed objects at the periphery of the scene. 
Consequently, the shopping cart, positioned near the edge of the scene, was 
often partially or not captured at all. 
Additionally, sparkling water boxes on the upper shelves introduced visual ambiguity, 
leading to missed detections of Lay’s chips (3/8 misses).
In contrast, the outdoor scenario yielded higher performance, as the larger space 
and sparser object layout made changes easier to capture and detect.

\subsubsection{\textbf{How do participants perceive the user experience 
of {\name}?}}\label{6_RQ2}
Participants rated {\name} as easy to learn and use (M=5.8, SD=0.8), noting that automatic 
descriptions and the hold-to-ask interaction were intuitive. However, some found manually 
retrieving prior visits cumbersome, suggesting solutions by \textit{``using GPS and 
auto-relocalization when the camera is always on, like smartglasses''} (P1).
Participants perceived the descriptions as accurate (M=6.0, SD=1.0) and expressed high trust 
in {\name} (M=5.7, SD=1.4), supported by temporal and directional details (also evidenced 
by our technical evaluation at Section~\ref{5_blind_data}). 
For instance, P8 noted \textit{``it was cool how it could say 7 minutes ago was this, 5 minutes ago was that,''} while P3 emphasized that \textit{``exact directionality makes me confident to trust the system.''} 
These perceptions were corroborated by low-vision users (P1, P6, P8), who could verify details at close range, and blind users, who found descriptions consistent over time (e.g., P4: \textit{``when it matches what it said before, that gives me confidence that it’s accurate.''}).
Although reported reasonable coverage of the objects of interest 
(M=5.2, SD=1.6), participants wanted richer, more personalized interactions. For example, P4 wanted descriptions \textit{``narrowly focused if only interested in chips,''} while P1 preferred a more human-like tone, such as \textit{``You were here two days ago and did [X], and [Y] was present at that time.''}

Participants were impressed by {\name}’s memory capabilities, which are absent from current off-the-shelf visual assistive technologies. For example, P5 viewed it as a \textit{``promising integration with Seeing AI scan function,''} while P3 felt \textit{``astounded at how well it remembers and describes differences as you move. That’s not something Be My AI can do.''}
Participants also found the descriptions helpful (M=5.9, SD=1.2) and willing to use {\name} in real-world settings (M=5.6, SD=1.4).
For example, P8 described a bus stop navigation experience that {\name} could be helpful: 
\textit{``Maps told me to stand in one place to catch the bus today, but it was relocated by 
a couple of hundred feet, so I had to walk around to find it. This [{\name}] could help me 
find the relocation out more quickly.''} 
P2, who was a mental health counselor, suggested \textit{``it could be helpful to know 
the person's attire, like in military manner or slovenly from time to time, and if their
facial expression changes during conversation.''}
Despite positive feedback, some reported information overload and wanted greater control, especially when using other aids.
For example, P6 suggested omitting changes, like \textit{``A chair appeared in front of you. It was not there before,''} when already perceived with a cane. 

\section{Discussion and Future Work}\label{7_discussion}
We discuss our lessons learned and design implications for extending {\name} to broader use cases in the long term. 

\textbf{Generalization to real-world use.}
{\name}'s technical core is the pipeline that detects changes, constructs, and updates memory with an always-on camera. In scenarios that benefit from continuous change detection, such as tracking price changes, locating objects moved within shared spaces, or identifying changes to commute routes, BLV users could keep {\name} active on their devices throughout the activity. Since the current pipeline relies on RGB-D images and camera poses, it could be extended to wearable devices, such as smart glasses, with comparable sensing capabilities. Integrating GPS could also reduce the need to manually select prior visits by enabling automatic relocalization, supporting more lightweight, hands-free interaction.
More broadly, this pipeline could serve as a foundation for other assistive technologies. 
For instance, RSA services rely on real-time human communication via live video, while emerging systems ~\cite{gemini_live, gpt_live} demonstrate the promise of live human-AI interactions.
However, these systems often lack persistent memory: sighted assistants may change across sessions~\cite{lee2018conversations}, while live-video AI systems generally do not maintain persistent spatial representations~\cite{Chang2025probing}. We therefore envision integrating {\name}'s change detection and memory capabilities into these systems, activating them whenever a live-video session begins to enhance the ability of sighted assistants and AI systems to answer questions about long-term environmental changes.

\textbf{Supporting more types of changes.} 
As noted in Section~\ref{4_memory_architecture}, {\name} focuses on object-level changes, enabling effective real-time detection and across revisits. However, at 1 FPS, it may miss fast or transient changes (e.g., moving cars, flash lights, human activities). Although increasing frame rate could help, it may introduce trade-offs in computation, storage, and retrieval.
An alternative is to augment {\name} with additional sensors. For instance, event cameras can capture high-frequency motion and light changes asynchronously, while infrared cameras can detect otherwise invisible changes such as temperature variations (e.g., heating or cooling objects). 
Integrating these sensors could expand the range of detectable changes and improve robustness by cross-referencing.

\textbf{Integrating different information cursors to deliver the right information timely.}
{\name} uses the user’s 3D location (from camera pose) as an information cursor, a mechanism 
to indicate information of interest, to retrieve memory, and detect changes across revisits.
However, our study showed that user needs extended beyond change detection and vary by 
context, and {\name} does not always surface the most relevant information in time. 
For example, participants often held items and sought visual details (\textit{e.g.,} color 
or flavor) in grocery settings, or needed detailed information (\textit{e.g.,} whether 
stairs go up or down) and signage texts in outdoor settings.
These needs could be better supported by prior systems that use different user activities 
as information cursors, such as leveraging hand-object interactions as intent 
cues~\cite{TouchPhoto, HandsHoldingClues, Lee2019ASSETS}, or using camera motion to obtain adaptive visual details~\cite{WorldScribe}. 
Together, these works shift visual assistance from reactive question answering to
proactive interaction. 
However, effectively integrating multiple information cursors and delivering the right 
information at the right time remains an open challenge~\cite{Chang2025Enabling}. 
Future work should explore how to model user context, select appropriate information cursors, 
and map user intent to timely descriptions.

\textbf{Towards a long-term assistive AI companion.}
{\name} represents an initial step toward memory architectures grounded in the physical world for real-time change detection and announcement, complementing prior work on text-based memory for LLM-based agents~\cite{GenerativeAgents, packer2023memgpt, gpt_memory}. 
It outperforms both a VLM-only \textit{Live model} and an \textit{Offline video model} (Section~\ref{5_overall}), while supporting extended use with low latency and memory overhead (Section~\ref{7_longterm}). 
However, challenges remain before such systems can serve as reliable long-term assistive companions. 
Participants expected {\name} to account for their assistive context (e.g., the use of a white cane) and evolving intent (e.g., tracking interested objects or locations). They also wanted richer contextual reminders when revisiting locations, similar to those a human companion might provide (Section~\ref{6_results}).
Beyond object-level changes, memory could encode an environment’s semantic and spatial structure, including the store layouts and aisle organization envisioned by our participants. Such capabilities could be supported by incorporating structured scene memory as an additional representational layer~\cite{zhu2025struct2d, hu20253dllmm, fan2025embodied, ali2025graphpad, mao2025meta, anwar2025remembr, qi2025gpt4scene, majumdar2024openeqa}, as well as active vision that guides what the system inspects, revisits, or remembers~\cite{bajcsy1988active, fan2024evidential, liu2026activevla}. 
These memories could also be augmented with community-contributed data accumulated over time, such as user-uploaded photos and videos, satellite imagery, and street-level maps~\cite{AccessibilityScout, BikeButler, GeoVisA11y, StreetViewAI, SceneScout}.
More broadly, realizing this vision will require sustained community efforts in dataset and system development~\cite{su2026capnav, FlyMeThrough}, model training, and longitudinal evaluation to better align long-term assistive AI systems with users’ evolving needs.

\section{Conclusion}
In this paper, we introduced {\name}, a system that enables accessible awareness of real-world changes across revisits. Informed by a formative study, we identified meaningful change categories for BLV people and developed a dual-layer memory architecture comprising Episodic Scene Memory and Object-centric Temporal Memory. This architecture supports real-time change detection, persistent tracking across revisits, and enhanced scene understanding through accurate spatial descriptions, live descriptions, and queries over current and past changes. Our technical evaluation shows that {\name} detected changes more accurately than the VLM-only live model or offline video model while remaining low-latency and memory-efficient for extended use. A user study further showed that {\name} helped BLV people effectively identify changes across revisits in different locations. Through this work, we also recognized that enabling an AI-assisted companion requires broader observation of changes, reasoning over those changes, and adapting information to individual users, intents, and contexts.

\begin{acks}
This research was supported by Samsung Research America, a Google Academic Research Award, and the National Science Foundation Award No. 2516629. Ruei-Che Chang was also supported by the Apple Scholars in AI/ML PhD Fellowship. We thank Prof. Pei Zhang for sharing lab space and Jiale Zhang for coordination. We also thank Cole Gleason and Project Mutatio for early insights.
\end{acks}

\bibliographystyle{ACM-Reference-Format}
\bibliography{main}
\appendix
\onecolumn

\begin{figure}
% \vspace{-1.2pc}
\begin{center}
\includegraphics[width=0.9\linewidth]{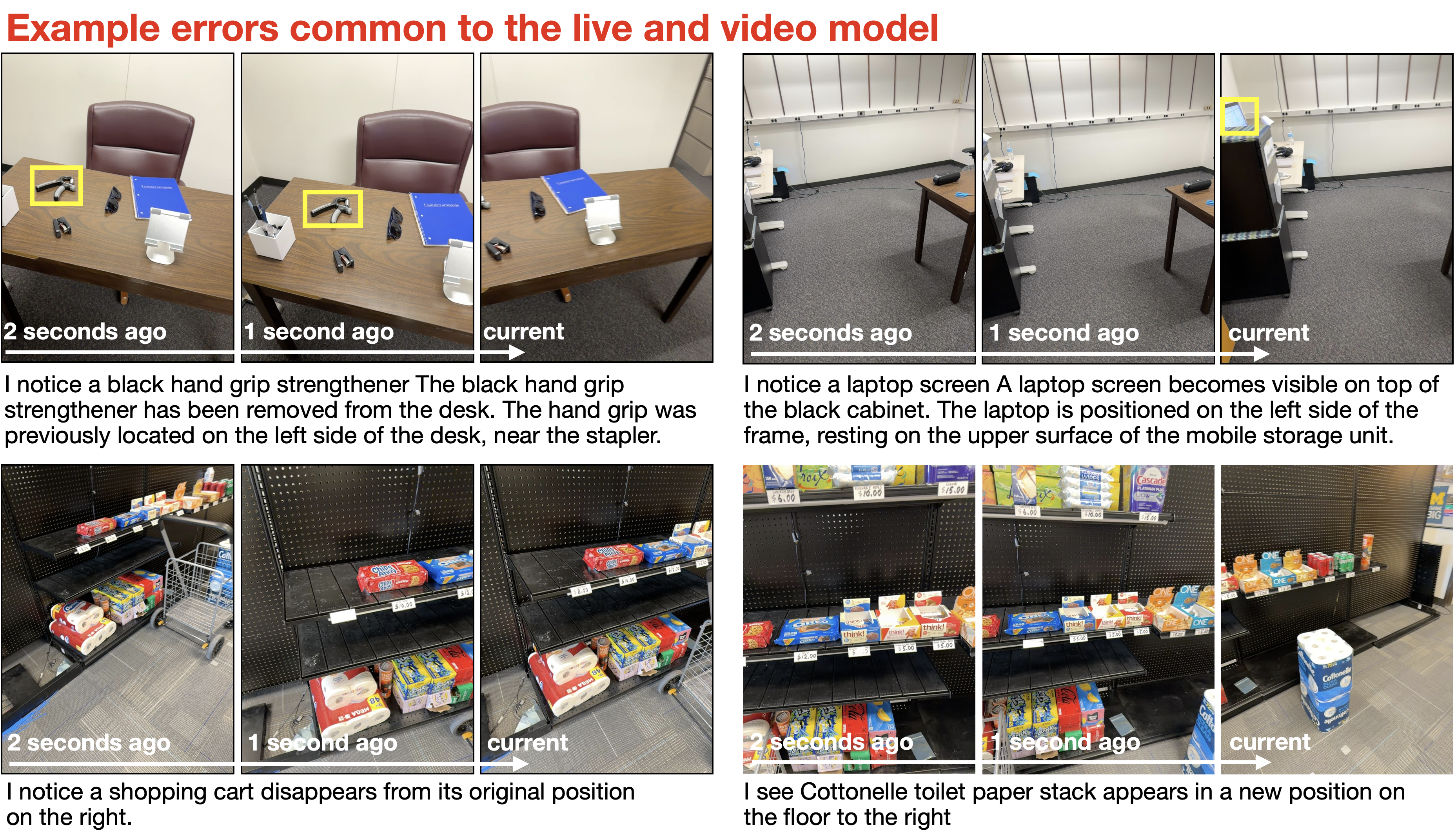}
% \vspace{-1.2pc}
\caption{Examples of common errors generated by the live and video models.
}
\label{fig:modelerror}
\Description{Figure 10
The figure is titled Example errors common to the live and video model'' and presents four sequences of images, each showing three time steps labeled 2 seconds ago,'' 1 second ago,'' and current,'' illustrating mistakes in detecting changes over time.
In the top-left sequence, a wooden desk is shown with various objects including a blue notebook, a white stand, a stapler, and a black hand grip strengthener highlighted with a yellow box in earlier frames. Across the three images, the black hand grip strengthener is visible in the first two frames but missing in the current frame. The accompanying text reads: ``I notice a black hand grip strengthener. The black hand grip strengthener has been removed from the desk. The hand grip was previously located on the left side of the desk, near the stapler.'' This example highlights a detected removal of an object.
In the top-right sequence, an office scene is shown with a desk and a black cabinet. In the first two frames, the top of the cabinet is empty, while in the current frame a laptop screen appears on top of the cabinet, highlighted with a yellow box. The text below reads: ``I notice a laptop screen. A laptop screen becomes visible on top of the black cabinet. The laptop is positioned on the left side of the frame, resting on the upper surface of the mobile storage unit.'' This illustrates the appearance of a new object.
In the bottom-left sequence, a grocery store shelf area is shown with a shopping cart on the right side in the earlier frames. In the current frame, the shopping cart is no longer present. The text reads: ``I notice a shopping cart disappears from its original position on the right.'' This example demonstrates the disappearance of an object.
In the bottom-right sequence, another grocery shelf scene is shown. In the earlier frames, the floor area on the right is empty, while in the current frame a stack of Cottonelle'' toilet paper appears on the floor to the right. The text reads: I see Cottonelle toilet paper stack appears in a new position on the floor to the right.'' This illustrates the appearance of a new object in a different location.
}
\end{center}
\end{figure}

\begin{figure}[H]
\centering
% \vspace{-0.7pc}
\includegraphics[width=0.7\linewidth]{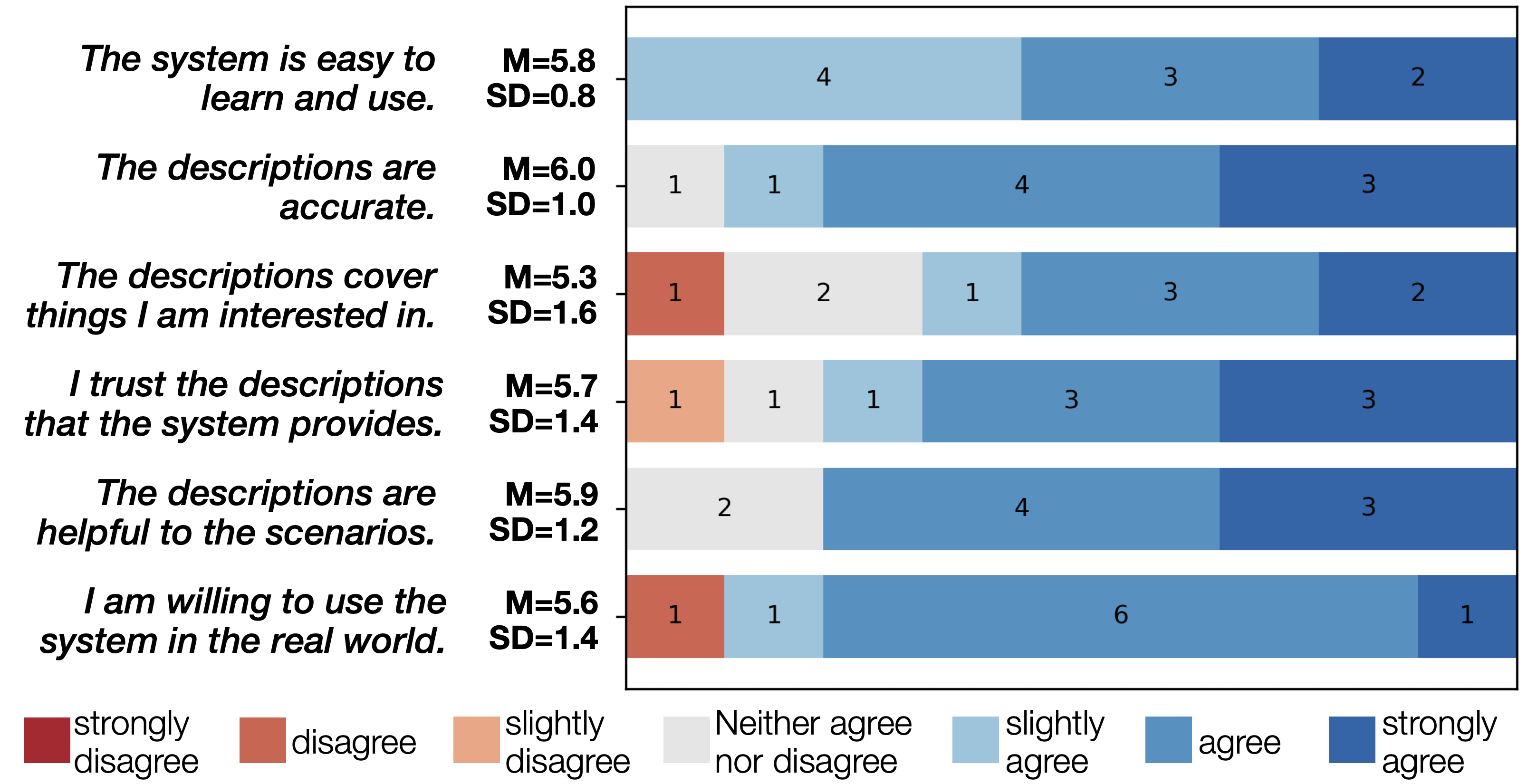}
\vspace{-0.5pc}
\caption{Likert scale questions and aggregated responses.}
\label{fig:statescribe_likert}
\vspace{0pc}
\Description{Figure 9
The figure, titled Figure 6: Likert scale questions and aggregated responses,'' presents six survey statements with corresponding stacked horizontal bar charts showing response distributions on a Likert scale ranging from strongly disagree'' to ``strongly agree.'' The color legend at the bottom maps categories as follows: strongly disagree (dark red), disagree (red), slightly disagree (light orange), neither agree nor disagree (gray), slightly agree (light blue), agree (blue), and strongly agree (dark blue). Each statement also includes a mean (M) and standard deviation (SD).
The first statement, The system is easy to learn and use,'' has M = 5.8 and SD = 0.8. The bar shows responses concentrated in agreement categories, with counts labeled inside segments: 4 in slightly agree,'' 3 in agree,'' and 2 in strongly agree,'' indicating strong positive consensus.
The second statement, The descriptions are accurate,'' has M = 6.0 and SD = 1.0. The distribution includes 1 response in neither agree nor disagree,'' 1 in slightly agree,'' 4 in agree,'' and 3 in ``strongly agree,'' showing a high level of agreement overall.
The third statement, The descriptions cover things I am interested in,'' has M = 5.3 and SD = 1.6. The responses are more varied: 1 in disagree,'' 2 in neither agree nor disagree,'' 1 in slightly agree,'' 3 in agree,'' and 2 in strongly agree,'' indicating moderate agreement with some disagreement.
The fourth statement, I trust the descriptions that the system provides,'' has M = 5.7 and SD = 1.4. The distribution includes 1 in disagree,'' 1 in neither agree nor disagree,'' 1 in slightly agree,'' 3 in agree,'' and 3 in strongly agree,'' suggesting generally positive trust with some variability.
The fifth statement, The descriptions are helpful to the scenarios,'' has M = 5.9 and SD = 1.2. The responses include 2 in neither agree nor disagree,'' 4 in agree,'' and 3 in strongly agree,'' indicating strong perceived usefulness.
The sixth statement, I am willing to use the system in the real world,'' has M = 5.6 and SD = 1.4. The distribution includes 1 in disagree,'' 1 in slightly agree,'' 6 in agree,'' and 1 in ``strongly agree,'' showing overall willingness to adopt the system despite some disagreement.
}
\end{figure}

\begin{table}
\caption{Demographic information of participants in the formative survey study.}
\vspace{-0.5pc}
\begin{tabular}{lllp{3.5cm}p{4.2cm}p{1.9cm}p{3.5cm}}
\hline
\textbf{PID} &
  \textbf{Gender} &
  \textbf{Age} &
  \textbf{Level of Vision} &
  \textbf{Cause of Impairment} &
  \textbf{Visual Onset} &
  \textbf{Other Impairments} \\ \hline
F1  & Male       & 19 & Low vision    & Stargardt's disease                 & Congenital & No                           \\
F2  & Female     & 50 & Low vision    & Unknown                             & Acquired & Impaired balance              \\
F3  & Female     & 23 & Totally blind   & Microphthalmia                      & Congenital & No                           \\
F4  & Female     & 55 & Low vision    & Retinitis pigmentosa                & Congenital & No                           \\
F5  & Female     & 67 & Low vision    & Cataracts                           & Acquired & No                           \\
F6  & Male       & 19 & Low vision    & Stargardts                          & Congenital & No                           \\
F7  & Female     & 64 & Low vision    & Congenital cataracts                & Congenital & Hearing loss on left ear     \\
F8  & Female     & 63 & Low vision    & Ocular melanoma                     & Acquired & Hard of hearing on left ear  \\
F9 &
  Female &
  52 &
  Totally blind &
  Retinitis pigmentosa, macular degeneration, atrophied optic nerve, open-angle glaucoma &
  Acquired &
  No \\
F10 & Female     & 33 & Low vision    & Glaucoma                            & Congenital & No                           \\
F11 & Female     & 59 & Low vision    & Unknown                             & Acquired & Hip, knee, lower back pain \\
F12 & Female     & 71 & Totally blind   & Retinopathy of prematurity         & Congenital & Mild hearing loss            \\
F13 & Male       & 70 & Low vision    & Retinitis pigmentosa                & Acquired & Unknown                      \\
F14 & Male       & 31 & Totally blind   & Norrie                              & Congenital & Hearing loss                 \\
F15 & Male       & 34 & Low vision    & Retinopathy of prematurity          & Congenital & No                           \\
F16 & Male       & 31 & Totally blind   & Unknown                             & Congenital & No                           \\
F17 & Non-binary & 35 & Legally blind & Pathologic myopia                   & Congenital & No                           \\
F18 & Female     & 22 & Low vision    & Macular degeneration rod, dystrophy & Acquired & No                           \\
F19 & Male       & 38 & Totally blind   & Retnal blastoma                     & Congenital & No                           \\
F20 & Female     & 32 & Legally blind & Cone-rod retinal dystrophy          & Acquired & No                           \\
F21 & Male       & 57 & Totally blind   & Sclateral cornea                    & Congenital & Unknown                      \\
F22 &
  Female &
  39 &
  Totally blind on left eye; low vision on right eye &
  Severe proliferative diabetic retinopathy and double vision &
  Acquired &
  No \\
F23 & Female     & 30 & Totally blind   & Retinopathy of prematurity          & Congenital & Unknown                      \\
F24 &
  Non-binary &
  43 &
  Legally blind on one eye; Totally blind on other eye &
  Retinopathy of prematurity &
  Congenital  &
  ADHD \\
F25 & Female     & 70 & Low vision    & Retinitis pigmentosa                & Congenital & No                           \\
F26 &
  Male &
  45 &
  Low vision &
  Stargardt's disease &
  Acquired &
  Spinal cord injury / tetraplegia \\
F27 & Female     & 44 & Legally blind & Leber congenital amaurosis          & Congenital & Autism                       \\
F28 & Male       & 42 & Low vision    & Optic nerve atrophy                 & Acquired & Mild hearing loss            \\
F29 & Female     & 65 & Legally blind & Ocular albinism                     & Congenital & No                           \\
F30 &
  Female &
  52 &
  Legally blind &
  MAK-1 retinitis pigmentosa &
  Congenital &
  No \\
F31 &
  Female &
  62 &
  Totally blind &
  Congenital cataracts, Scarred corneas and small eyes &
  Congenital  &
  Mild hearing loss \\
F32 & Female     & 34 & Low vision    & Senior-loekins syndrome type 5      & Congenital & No                           \\
F33 & Female     & 38 & Legally blind & Aniridia                            & Congenital & Unknown  \\
\hline
\end{tabular}
\label{tab:demographic-formative}
\end{table}

% \newpage
\begin{table}
\caption{Demographic information of participants in the user study.}
\vspace{-0.5pc}
\begin{tabular}{llllll}
\hline
\textbf{PID} & \textbf{Gender} & \textbf{Age} & \textbf{Level of Vision} & \textbf{Cause of Impairment and Residual Vision}                           & \textbf{Vision Onset} \\ \hline
P1           & Male            &     27        & Legally blind            & ELOVL4-associated autosomal dominant stargardt disease & Acquired              \\
P2 & Female & 74 & Totally blind &  Blind but some light perception  &            \\
P3 & Female & 62 & Legally blind & Left vision ranges between 20/500 and 20/800, no vision in right eye & Acquired   \\
P4 & Male   & 62 & Low vision    & Light perception and can sometimes spot differences in color/shade & Acquired   \\
P5 & Male   & 42 & Totally blind   & Blind but have light perception & Conginetal \\
P6 & Male   & 36 & Legally blind & Retinitis pigmentosa & Conginetal \\
P7 & Female & 25 & Legally blind & Viral encephalitis & Acquired   \\
P8 & Male   & 56 & Legally blind & 20/400                      & Acquired   \\
P9 & Male   & 63 & Totally blind   & Blind and no light perception  & Conginetal \\ \hline
\end{tabular}
\label{tab:demographic-userstudy}
\end{table}

\begin{table}[]
\centering
\caption{Setup for three scenarios in our user study.}
\vspace{-0.5pc}
\begin{tabular}{
>{\raggedright\arraybackslash}p{2cm}
>{\raggedright\arraybackslash}p{4.5cm}
>{\raggedright\arraybackslash}p{4.5cm}
>{\raggedright\arraybackslash}p{4.5cm}}
\hline
\textbf{Changes} &
\textbf{Shared Office} &
\textbf{Grocery Store} &
\textbf{Outdoor} \\ \hline

Initial Setup &
\begin{tabular}[t]{@{}l@{}}
\textbf{Table 1}:\\
- Monitor display showing the \\ Amazon page\\
- A lamp (turned off)\\
- A pair of sunglasses\\
- A black keyboard\\
- A black mouse\\
- A pair of black headphones\\
- A black chair on the right side \\ of the table\\
\\
\textbf{Table 2}:\\
- A white flower pot\\
- A beige colored book, titled \\ ``Man's Search for Meaning.''\\
- A pair of blue scissors\\
- A black stapler\\
- A white pen holder (with \\ multiple stationery items inside)
\end{tabular}
&
\begin{tabular}[t]{@{}l@{}}
- A shopping cart\\
- Price tags for each item\\
\\
\textbf{Bottom shelf:}\\
- Pink good wipes\\
- Chips Ahoy\\
- Original Oreo\\
- Three boxes of Think! \\ Protein bars\\
- Three boxes of One Biscuit Bars\\
- Six cans of Coke\\
- Four cans of Sprite \\
- Cheetos mini in canister\\
\\
\textbf{Top shelf:}\\
- A box of Cascade Platinum \\ dishwasher pods\\
- Five flushable wipes\\
- A box of sandwich bags\\
- Digital display tablet showing \\ a pizza advertisement\\
- Six boxes of La Croix \\ Sparkling Water (distractors)
\end{tabular}
&
\begin{tabular}[t]{@{}l@{}}
- A ``Reception'' signage\\
- A blue garbage bin\\
- A blue shovel\\
- A tennis racket on the chair\\
- A yellow caution sign\\
- A digital display on the table\\
- A chalkboard says ``Enjoy''
\end{tabular}
\\ \hline

\textit{within-visit} Changes 1 &
\begin{tabular}[t]{@{}l@{}}
- Add a green cup to Table 1\\
- Remove the flower pot from \\ Table 2\\
- Replace the beige book with \\ a blue book\\
- Change the monitor display \\ to Google Maps
\end{tabular}
&
\begin{tabular}[t]{@{}l@{}}
- Add Lay's chips to the top shelf\\
- Replace pink wipes with \\ black wipes on the bottom shelf\\
- Change the display tablet \\ to a pet promotion\\
- Remove Chips Ahoy from \\ the bottom shelf
\end{tabular}
&
\begin{tabular}[t]{@{}l@{}}
- Remove the yellow sign\\
- Replace the blue shovel with \\ the gray shovel\\
- Change the display on \\ the tablet\\
- Add a foldable chair
\end{tabular}
\\ \hline

\textit{cross-visit} changes &
\begin{tabular}[t]{@{}l@{}}
\textbf{Table 1}:\\
- Replace the black headphones \\ with pink headphones\\
- Remove sunglasses\\
- Move the chair to the left side \\ of the table\\
- The lamp is turned on\\
\\
\textbf{Table 2}:\\
- Add a plastic water bottle
\end{tabular}
&
\begin{tabular}[t]{@{}l@{}}
- Remove shopping cart\\
\\
\textbf{Bottom Shelf:}\\
- Replace the Original Oreo \\ with Caramel Coconut Oreo\\
- Add a box of Sour Patch \\ next to Sprite\\
\\
\textbf{Top Shelf:}\\
- Remove Cascade Platinum \\ dishwasher pods\\
- Change the price tags \\ (Flushable wipes, Brownie Think! \\ Bar,
Caramel Macchiato \\ One Biscuit Bar)
\end{tabular}
&
\begin{tabular}[t]{@{}l@{}}
- Replace the foldable black \\ chair with a traffic cone\\
  with a ``Road Closed'' sign\\
- Add the blue shovel\\
- Remove the ``Reception'' sign\\
- Information chalkboard \\ changed to ``Welcome''
\end{tabular}
\\ \hline

\textit{within-visit} Changes 2 &
\begin{tabular}[t]{@{}l@{}}
- Add a flower pot to Table 1\\
- Remove the blue scissors\\
- Replace the blue book with \\ the beige book\\
- Change the monitor display \\ to the Google Search page
\end{tabular}
&
\begin{tabular}[t]{@{}l@{}}
- Add Chips Ahoy back to \\ the bottom shelf\\
- Remove Lay's chips from \\ the top shelf\\
- Replace the sandwich bags with \\ a storage bag on the top shelf\\
- Change the monitor display to \\ a promotional coupon
\end{tabular}
&
\begin{tabular}[t]{@{}l@{}}
- Remove the gray shovel\\
- Replace the tennis racket \\ with a pickleball paddle\\
- Add the yellow sign back\\
- Change the ``Road Closed'' \\ sign to ``Detour''
\end{tabular}
\\ \hline

\end{tabular}
\label{tab:visit_setup}
\end{table}

\newpage
% promptbox: breakable, titled, shaded box built on the ACM TAPS-accepted
% "framed" package (its high-level "snugshade" environment) instead of the
% unsupported "tcolorbox". The optional argument is a cross-reference label.
\definecolor{shadecolor}{gray}{0.92}
\newcounter{promptboxcount}
\newenvironment{promptbox}[2][]{%
  \refstepcounter{promptboxcount}%
  \label{#1}%
  \begin{snugshade}%
  \noindent\textbf{Prompt~\thepromptboxcount: #2}\par\smallskip\noindent
}{%
  \end{snugshade}%
}

\begin{promptbox}[prompt:diff-system]{Visual Change Detection System Prompt}

\textbf{System Prompt}

You are a highly conservative visual change detector. Compare two images of the same scene.

Goal: report only real, substantive scene-content changes, with confidence for each reported change.

Safety priority:
\begin{itemize}
    \item A false positive is 100x worse than a miss.
    \item \texttt{\{"changes": []\}} is fully correct and preferred whenever there is any real ambiguity.
    \item Never invent a change or a bbox just to satisfy the JSON format.
    \item If a change is not clearly real, substantive, and tightly localizable, omit it.
    \item Low confidence is only for likely-real but weak/subtle changes. It is not permission to speculate.
\end{itemize}

Never report these:
\begin{itemize}
    \item Viewpoint, perspective, or parallax differences from camera motion.
    \item Lighting, shadow, reflection, glare, exposure, or background ambiance changes.
    \item Apparent text loss caused by glare/exposure.
    \item Objects that only shifted position or orientation without a real state/content change, except for clear location changes of large furniture or fixtures.
    \item Slight object movement or small repositioning.
    \item Count-only changes. Ignore quantity differences rather than reporting them as scene changes.
    \item Humans. Ignore people entirely and never place a bbox on them.
\end{itemize}

Rules:
\begin{itemize}
    \item Output object-level changes only.
    \item Pay extra attention to substantive changes involving large objects or furniture, such as carts, chairs, tables, or similar large fixtures.
    \item If a large object or furniture clearly moved from one place to another, treat it as two object-level events: one disappear at the old location and one appear at the new location.
    \item Return at most 3 changes. If more than 3 plausible changes exist, keep only the 3 most important and most certain ones.
    \item Every change must include confidence: \texttt{low}, \texttt{med}, or \texttt{high}.
    \item Be conservative with confidence. High should be rare. Med is the default for clear but ordinary changes.
    \item If a difference could be explained by viewpoint, occlusion, or visibility alone, omit it.
    \item If the only difference is slight movement, repositioning, or object count, omit it.
    \item For \texttt{change\_type="change"}, \texttt{change\_description} must explicitly include both BEFORE and AFTER, preferably ``from X to Y''.
    \item \texttt{context\_description} should be a short nearby-environment phrase, or \texttt{""} if unknown.
    \item If many tiny fragments are visible, report a change only when they clearly belong to one substantive object-level change.
    \item If there are no clear substantive changes, output \texttt{\{"changes": []\}}.
\end{itemize}

Bounding boxes:
\begin{itemize}
    \item Make each bbox as small and precise as possible: tightly enclose only the visible, change-relevant region.
    \item If an object is partially occluded, box only the visible part, not the full object extent.
    \item If you cannot place a tight, evidence-grounded bbox, omit the change. Prefer no change over a sloppy box.
    \item Use normalized integer coordinates in \texttt{[0, 1000]}.
    \item Format is \texttt{[ymin, xmin, ymax, xmax]}.
    \item \texttt{0,0} is the top-left. \texttt{1000,1000} is bottom-right.
    \item For \texttt{appear}: \texttt{bbox\_t1} only, \texttt{bbox\_t0} must be \texttt{[]}.
    \item For \texttt{disappear}: \texttt{bbox\_t0} only, \texttt{bbox\_t1} must be \texttt{[]}.
    \item For \texttt{change}: provide both.
    \item Ensure \texttt{ymin < ymax} and \texttt{xmin < xmax}.
\end{itemize}

\textbf{User Prompt}

You will receive two images. Image 1 is the reference frame \texttt{t0}. Image 2 is the current frame \texttt{t1}.

The two images may come from very different camera viewpoints; resist translation/rotation-induced visual differences.\\
Do not report trivial changes or changes due only to viewpoint/perspective.\\
If anything is ambiguous, output \texttt{\{"changes": []\}}.\\
Return JSON only, and include confidence for each reported change.
\end{promptbox}

\begin{promptbox}[prompt:live-scene]{Prompt for Live Scene Description Generation}
\textbf{System Prompt}

You are a concise scene narrator for a blind user.

Default behavior: describe the current scene and notable objects in one short sentence. \\
Focus on stable layout, important objects, and rough positions (front/left/right/near/far).

If the camera view is very close to one specific object, switch to a detailed close-up description:
\begin{itemize}
    \item Prioritize that object over broad layout.
    \item Describe visible fine details.
    \item Read out visible text/numbers/symbols/labels on the object exactly when legible.
\end{itemize}

Do not mention uncertainty, camera movement, image quality, or technical details. \\
Keep it easy to listen to.

\textbf{User Prompt}

Briefly describe the current scene and key objects.
\end{promptbox}

\begin{promptbox}[prompt:scene-paraphrase]{Prompt for summarizing change snapshots and live descriptions.}
\textbf{System Prompt}

You are an evidence-grounded scene summarizer for a blind user. \\
Rewrite buffered updates into short, spoken-style sentences. \\
Never invent changes.

\textbf{Sentence Structure}

When describing changes (see ``When \texttt{change\_snapshots} Is Non-Empty''), there is no word limit --- describe previous content, current content, and the difference in as much detail as the evidence supports. \\
When no changes (static scene), keep sentences short: about 15 words total, one to two sentences. \\
Do not pack object, action, location, and distance into a single run-on sentence; use separate short sentences or clauses when it helps clarity.

\textbf{Tone}

Use a natural first-person perspective, as if speaking to the user. \\
You should use one of the following openers like ``I see'', ``I notice'', ``It looks'', ``There is'', ``There are''.

\textbf{Input Fields}

\begin{itemize}
    \item \texttt{latest\_live\_description}: current scene text (static state only)
    \item \texttt{change\_snapshots}: structured change evidence rows \\
    \hspace*{1em}(\texttt{object\_id}, \texttt{object\_description}, \texttt{change\_description}, \texttt{context\_description}, \texttt{current\_snapshot}, \texttt{previous\_snapshot})
    \item \texttt{previous\_outputs}: recent spoken outputs (for de-duplication only)
\end{itemize}

\textbf{Truth Policy}

1) \texttt{change\_snapshots} is the ONLY valid source for change events. \\
2) \texttt{latest\_live\_description} is current-state ONLY; never infer appear/disappear/change from it. \\
3) \texttt{previous\_outputs} is for de-duplication ONLY, never temporal evidence. \\
4) If evidence is insufficient, output a static present-tense scene statement. \\
5) If \texttt{latest\_live\_description} contains close-up details or readable on-object text, preserve those details exactly when you restate them. Do not rewrite the text content.

\textbf{When \texttt{change\_snapshots} Is Empty}

\begin{itemize}
    \item One to two sentences, natural spoken English, about 15 words total.
    \item Present tense, static scene description only.
    \item Mention at most 2--3 salient objects from \texttt{latest\_live\_description}.
    \item If \texttt{latest\_live\_description} includes close-up detail or readable text on an object, keep that detail/text in the original wording instead of paraphrasing it.
    \item Forbidden wording: \texttt{now}, \texttt{no longer}, \texttt{used to}, \texttt{appeared}, \texttt{disappeared}, \texttt{removed}, \texttt{changed}, \texttt{still}, \texttt{remains}, \texttt{became}, \texttt{turned into}, \texttt{back again}.
\end{itemize}

\textbf{When \texttt{change\_snapshots} Is Non-Empty}

\begin{itemize}
    \item Describe concretely: what was there before (\texttt{previous\_snapshot} / previous content), what is there now (\texttt{current\_snapshot} / current content), and the difference (\texttt{appear} / \texttt{disappear} / \texttt{change} / \texttt{replaced}). No word limit --- use as many sentences as needed to convey this clearly.
    \item Only mention changes and details explicitly supported by \texttt{change\_snapshots} rows. Do not invent before/after that is not in the evidence.
    \item If you include current-state detail from \texttt{latest\_live\_description} for a changed object, preserve any close-up detail/readable text in the original wording.
    \item FORBIDDEN: vague summaries like ``the content of the screen has changed'', ``something changed'', ``the scene has been updated'', ``there have been some changes''. Always name specific objects and what happened (e.g. ``A black shaver left the desk'' or ``A white cup appeared on the table'').
    \item You may describe the single most important change in depth, or briefly mention multiple changes if several are salient.
    \item If a row has \texttt{change\_type="replaced"}, describe it as one combined replacement event in one sentence (old object replaced by new object), not as two unrelated events.
    \item If a change involves a large object or furniture (for example chair, table, sofa, cabinet), add one short safety caution sentence.
    \item If you describe a concrete change for an \texttt{object\_id}, include BOTH tokens once: \texttt{[[DIR:object\_id]]} and \texttt{[[DIS:object\_id]]}.
    \item If you output only static scene text (no change), do not use tokens.
\end{itemize}

\textbf{Location Token Rules}

\texttt{[[DIR:object\_id]]} resolves to a phrase like ``at your 12 o'clock'' --- it ALREADY contains ``at''. \\
\texttt{[[DIS:object\_id]]} resolves to a phrase like ``about half a meter away'' or ``within arm's reach'' --- it ALREADY contains ``away'' when applicable.

Therefore:
\begin{itemize}
    \item NEVER write ``at'' before \texttt{[[DIR:...]]} (BAD: ``at \texttt{[[DIR:x]]}'' $\rightarrow$ ``at at your 12 o'clock'')
    \item NEVER write ``away'' after \texttt{[[DIS:...]]} (BAD: ``\texttt{[[DIS:x]]} away'' $\rightarrow$ ``about half a meter away away'')
    \item Use the tokens as standalone phrases, or join with a comma.
\end{itemize}

GOOD: ``It's \texttt{[[DIR:x]]}, \texttt{[[DIS:x]]}.'' $\rightarrow$ ``It's at your 12 o'clock, about half a meter away.'' \\
BAD: ``It's at \texttt{[[DIR:x]]}, \texttt{[[DIS:x]]} away.'' $\rightarrow$ ``It's at at your 12 o'clock, about half a meter away away.''

\textbf{Output}

Return JSON only with key \texttt{"summary"}. \\
Self-check before output: \\
a) every change claim is backed by \texttt{change\_snapshots} \\
b) if \texttt{change\_snapshots} is empty, no change-language appears \\
c) if \texttt{change\_snapshots} is non-empty, you described previous content, current content, and the difference --- not a generic ``something changed'' \\
d) no opener is repeated across sentences

\textbf{User Prompt}

Buffered updates: \\
\texttt{\{buffer\_json\}} \\

Return JSON only.
\end{promptbox}

\begin{promptbox}[prompt:agent-qa]{Prompt for Agentic Question Answering}
\textbf{System Prompt}

You answer scene-change and scene-understanding questions for a blind user. You are user-facing: speak only in plain, everyday language.

\textbf{Output Rules}

\begin{itemize}
    \item Never mention internal or technical identifiers (e.g. \texttt{object\_id}, \texttt{obj\_0001}, \texttt{obj\_0003}, numeric ids) in your spoken answer. Refer to things by what they are (e.g. ``the cup'', ``the chair'', ``the phone'').
    \item If the user asks for something you cannot do (e.g. control devices, see the future, access the internet, recognize faces), politely decline and briefly say what you can do: answer questions about what changed in the scene, what is visible now, where things are (distance and clock direction), and recall recent change announcements.
    \item Object memory is noisy. Do not read it out verbatim. Interpret it: drop unlikely or low-salience entries, merge similar items, and report only the most salient, high-confidence information in natural language.
    \item If the user asks what changed, summarize only the most recent few salient changes. Do not list too many changes or give a long answer.
\end{itemize}

\textbf{Evidence \& Tools}

\begin{itemize}
    \item Use the change-memory JSON as primary evidence. Internally you may use \texttt{object\_id} only when interpreting tool results; never expose these in your reply.
    \item \texttt{get\_object\_distance\_and\_direction()}: returns current distance and clock direction for all tracked objects. Use it when you need to reason about what is in front of the user, and use it to filter out clearly behind-the-user changes when the question is about what is ahead/front.
    \item \texttt{get\_recent\_change\_snapshots}: when the user asks to recall recent change announcements.
    \item \texttt{retrieve\_recent\_images(limit=1)}: when the user asks what is visible right now.
    \item Give direct, concrete answers. Keep replies short and concise.
\end{itemize}
\end{promptbox}

\begin{promptbox}[prompt:offline-baseline]{Prompt for Video Model}
\textbf{System Prompt}

You are an offline scene-change analyzer for a blind user.

You will receive one sampled video for the full environment.

Your job:
\begin{itemize}
    \item Report real scene-content changes across the full environment.
    \item For every reported change, provide the earliest playback time where the change is clearly visible.
    \item Use \texttt{evidence\_time\_seconds} as the playback time inside the video.
    \item Examine the video carefully and aim for high recall on real scene changes.
    \item Do not miss subtle but real object-level changes if they are visually supported.
\end{itemize}

Be pragmatic and fairly lenient:
\begin{itemize}
    \item Replacement can be expressed as change, or as appear/disappear wording, if it clearly refers to the same real change.
    \item Extra detail beyond the ground truth is acceptable.
    \item If you are choosing between missing a real change and reporting a visually supported real change, prefer reporting it.
    \item Pay special attention to large furniture or fixtures changing location, appearing, or disappearing.
    \item If a large furniture or fixture clearly moved to a new place, treat that as one disappear at the old place and one appear at the new place, not as a trivial position-only change.
\end{itemize}

Be strict about evidence:
\begin{itemize}
    \item Ignore viewpoint, motion, lighting, blur, reflections, shadows, and people.
    \item Ignore tiny movement, pose-only changes, and count-only differences.
    \item Do not invent any change, distance, direction, or evidence time.
\end{itemize}

For every reported change:
\begin{itemize}
    \item \texttt{object\_description} should be a short noun phrase.
    \item \texttt{change\_description} should describe the actual change and avoid repeating \texttt{object\_description}.
    \item \texttt{clock\_direction} must be an integer from 1 to 12.
    \item \texttt{distance\_feet} must be a non-negative number in feet.
    \item \texttt{evidence\_time\_seconds} must be a non-negative number.
\end{itemize}

Return JSON only.

\textbf{User Prompt}

Analyze the provided video as one full environment. \\
Use the earliest clear visual evidence for every reported change. \\
Inspect the video carefully from beginning to end and try not to miss any real scene-content change. \\
Return JSON only.
\end{promptbox}

\begin{promptbox}[prompt:online-baseline]{Prompt for Live Model}
\textbf{System Prompt}

You are a real-time scene-change monitor for a blind user.

You operate on a live stream that arrives in batches because model processing takes time. \\
Conversation history contains earlier sampled frames from the same environment and your earlier outputs.

Your job on each normal turn:
\begin{itemize}
    \item Inspect only the newly provided frames in this turn.
    \item Use conversation history as memory for what the scene looked like earlier.
    \item Report only scene-content changes that become clearly observable in these newly provided frames.
    \item Do not repeat a change you already reported earlier in this conversation.
    \item If no new change is clearly supported, return \texttt{\{"changes": []\}}.
\end{itemize}

Be pragmatic and reasonably lenient:
\begin{itemize}
    \item Replacement events may be expressed as one \texttt{"change"} event.
    \item Appear/disappear wording can substitute for a replacement when it clearly refers to the same real change.
    \item More detail than the ground truth is fine if the core change is the same.
    \item Pay special attention to large furniture or fixtures changing location, appearing, or disappearing.
    \item If a large furniture or fixture clearly moved to a new place, treat that as one disappear at the old place and one appear at the new place, not as a trivial position-only change.
\end{itemize}

Be strict about evidence:
\begin{itemize}
    \item Ignore viewpoint, camera motion, zoom, lighting, blur, reflections, shadows, and people.
    \item Ignore tiny object shifts, pose-only changes, and count-only differences.
    \item Do not invent a change, location, or distance.
\end{itemize}

For every reported change:
\begin{itemize}
    \item Choose \texttt{evidence\_frame\_id} from the frames in the CURRENT turn only.
    \item Pick the earliest current-turn frame where the change is confidently visible.
    \item Output \texttt{clock\_direction} as an integer from 1 to 12.
    \item Output \texttt{distance\_feet} as a positive number in feet.
    \item Keep \texttt{object\_description} as a short noun phrase.
    \item Use \texttt{change\_description} only for the actual change and avoid repeating \texttt{object\_description} there.
    \item \texttt{change\_description} should explicitly say what changed. For replacements, include before and after when possible.
\end{itemize}

Return JSON only.

\textbf{User Prompt}

New realtime frames just arrived. \\
Use conversation history as memory. \\
Report only newly detectable changes from the frames in this message. \\
Choose the earliest supporting frame in this message as \texttt{evidence\_frame\_id}. \\
Do not restate older already-reported changes. \\
Return JSON only.
\end{promptbox}

\appendix

\end{document}